\documentclass[twocolumn,showpacs,preprintnumbers,amsmath,amssymb]{revtex4}

\usepackage{graphicx}
\usepackage{dcolumn}
\usepackage{bm}

\def\fun#1#2{\lower3.6pt\vbox{\baselineskip0pt\lineskip.9pt
\ialign{$\mathsurround=0pt#1\hfil##\hfil$\crcr#2\crcr\sim\crcr}}}
\def\lap{\mathrel{\mathpalette\fun <}}
\def\gap{\mathrel{\mathpalette\fun >}}

\def\mass{{\cal M}}

\def\msun{{\mass_\odot}}

\def\beq{\begin{equation}}
\def\eeq{\end{equation}}

\def\mh{M_{\bullet}}
\def\ms{m_{\star}}

\begin{document}


\title{Testing Properties of the Galactic Center Black Hole Using Stellar Orbits}
\author{David Merritt}
 \email{merritt@astro.rit.edu}
\affiliation{Department of Physics and Center for Computational Relativity and Gravitation, Rochester Institute of Technology, Rochester, NY 14623}
\author{Tal Alexander}
\email{tal.alexander@weizmann.ac.il}
\affiliation{Faculty of Physics,Weizmann Institute of Science, POB 26, Rehovot, Israel}
\author{Seppo Mikkola}
\email{mikkola@utu.fi}
\affiliation{Tuorla Observatory, University of Turku, V\"ais\"al\"antie 20, Piikki\"o, Finland}
\author{Clifford M. Will}
\email{cmw@wuphys.wustl.edu}
\affiliation{McDonnell Center for the Space Sciences, Department of Physics, Washington University, St. Louis, MO 63130}
\date{\today}

\begin{abstract}
The spin and quadrupole moment of the supermassive black hole at
the Galactic center can in principle be measured via astrometric monitoring
of stars orbiting at milliparsec (mpc) distances, allowing tests of
general relativistic ``no-hair''theorems \cite{Will-08}.
One complicating factor is the presence of perturbations from other
stars, which may induce orbital precession of
the same order of magnitude as that due to general relativistic effects.
The expected number of stars in this region is small enough that full 
$N$-body simulations can be carried out.
We present the results of a comprehensive set of such simulations,
which include a post-Newtonian treatment of spin-orbit effects.
A number of possible models for the distribution of stars and
stellar remnants are considered.
We find that stellar perturbations are likely to obscure the
signal due to frame-dragging for stars beyond $\sim 0.5$ mpc
from the black hole,
while measurement of the quadrupole moment is likely to require
observation of stars inside $\sim 0.2$ mpc.
A high fraction of stellar remnants, e.g. $10\msun$ black  holes,
in this region would make tests of GR problematic at all radii.
We discuss the possibility of separating the effects of stellar
perturbations from those due to GR.
\end{abstract}

\pacs{Valid PACS appear here}
\maketitle

\section{\label{sec:intro} Introduction}

The supermassive black hole (SBH) at the center of the Milky Way
galaxy is surrounded by a compact cluster of stars that
has been the target of observational surveys for more than a decade
\cite{Krabbe-95,Blum-96,Figer-00,Gezari-02,Paumard-06,Zhu-08}.
Near-infrared monitoring of stellar positions using adaptive optics 
techniques has allowed orbital reconstruction for roughly 30 stars at 
distances ranging from $10^0-10^2$ milliparsecs (mpc) from the SBH
\cite{Ghez-05,Ghez-08,Gillessen-09a}.
One of these stars (S2) has an orbital period of only $\sim 15$ yr
\cite{Schoedel-02,Ghez-03}
and its orbit has been followed for more than one full revolution;
astrometric data for S2 yield a well-constrained mass for the SBH,
$\mh = (3.95\pm 0.06)\times 10^6\msun$ 
(assuming a galactocentric distance of $8.0$ kpc) 
and a location on the plane
of the sky that is consistent with that of the radio source Sgr A$^*$
\cite{Reid-93,Eisenhauer-05,Ghez-08,Groen-08,Gillessen-09a,Gillessen-09b}.

The velocity of S2 near periapse is a few percent of the speed of light,
large enough that relativistic effects
like advance of the periastron become potentially measurable, 
even on time scales as short as a few years
\cite{Jaro-98,Fragile-00,Weinberg-05,Zucker-06}.
No such effects have so far been unambiguously observed \cite{Gillessen-09b}; 
one complicating factor
is the likely presence of a distributed mass (stars, stellar remnants,
dark matter etc.) within S2's orbit which could produce Newtonian
precession of the same order of magnitude as that due to
general relativity \cite{Rubilar-01,Zakharov-07}.

If the SBH is rotating, new phenomena occur for stars
orbiting at very small separations, $r\lap 1$ mpc.
Dragging of inertial frames and torques from the SBH's quadrupole 
moment $Q$ cause stellar orbital planes to precess, at rates that depend
respectively on the first and second powers of the hole's spin angular 
momentum $\mathbf{J}$.
These spin-related effects are small compared with in-plane
precession, but (in the absence of other non-spherically-symmetric
components of the gravitational potential) they contain
unambiguous information about $\mathbf{J}$ and $Q$.
In principle, observed changes in the orbital orientations
of just two stars would be sufficient to independently constrain
the four quantities $(\mathbf{J}, Q)$, allowing tests of general
relativistic ``no-hair'' theorems \cite{Will-08}.
The amplitude of these spin-related precessions is
very small, of order microarcseconds ($\mu$as) per year as seen from
the Earth.
Plans are being developed to achieve infrared astrometry at
this level \cite{GRAVITY-09,ASTRA-08}.

Such measurements will require the presence of at least a few
bright stars on mpc-scale orbits around the SBH.
While no such stars have yet been observed,
extrapolation of the observed stellar densities at distances of
$\sim 1$ pc from the SBH
suggests that of order $10^0-10^2$ stars should be present 
in this region.
Due to their finite numbers, these stars will generate
a non-spherically-symmetric component to the gravitational
potential with an amplitude that scales
as $\sqrt{N}\ms$, where $\ms$ is the mass of a typical star and $N$
is their number.
Simple arguments (\S\ref{sec:timescales}) suggest that such
stellar perturbations might produce changes in the orbital
orientations of test stars that are comparable in magnitude
to the spin-related effects.
This would complicate the testing of no-hair theorems by adding
what is effectively a source of noise to the measured
precessions.

Because the expected number of stars in this region is so small,
direct $N$-body integration of the equations of motion for all
$N$ stars is feasible.
The major technical requirements are a high degree of accuracy
in the $N$-body integrator and the inclusion of terms
describing the relativistic accelerations due to the SBH, 
including spinless, spin-orbit, and quadrupole-orbit contributions.

Here we present the results of a comprehensive set of such 
simulations.
Our primary goal is to evaluate the degree to which star-star perturbations
might obscure the signal due to the SBH's spin; hence we focus
on changes in orbital orientations rather than on the evolution of
the phase-space variables ($\mathbf{r},\mathbf{v}$) \cite{Kannan-09,Preto-09}.
We ignore all other systematic effects that might limit the
ability to carry out the high-precision astrometry for
stars in crowded fiels at the Galactic center
\cite{Weinberg-05,Fritz-09}.

In \S\ref{sec:timescales} we summarize the relevant time scales 
for orbital evolution near the Milky Way SBH.
\S\ref{sec:nbody} presents the post-Newtonian $N$-body equations of motion
including the lowest-order spin-orbit terms and describes the
$N$-body integrator.
Observational and theoretical constraints on the distribution of stars and 
stellar remnants near the Galactic center SBH are summarized in \S\ref{sec:models}, which
also describes the parametrized models used to construct the $N$-body
initial conditions. 
\S\ref{sec:results} summarizes the results from the integrations,
including estimates of the number of stars that can be
effectively used to measure $\mathbf{J}$ and $Q$.
\S\ref{sec:discuss} discusses how the presence of stellar perturbations
in the astrometric data can potentially be detected and removed
from the GR signal.
\S\ref{sec:conclude} sums up.

\section{\label{sec:timescales}Sources of orbital evolution}
\subsection{Basic quantities}

The orbital period of a star of semi-major axis $a$ orbiting
around the Milky Way SBH is
\beq
P = {2\pi a^{3/2}\over \sqrt{G\mh}}
\approx 1.48 {\tilde a}^{3/2}{\rm yr}
\label{eq:period}
\eeq
where $\mh$ is the mass of the SBH and 
$\tilde a$ is the star's semi-major axis in units 
of milli-parsecs (mpc). 
The second relation assumes $\mh=4.0\times 10^6\msun$ 
and $\ms\ll\mh$, assumptions which we adopt in the remainder of the paper.
The length scale associated with the event horizon of the SBH is
\beq
r_g \equiv {G\mh\over c^2} \approx 1.92\times 10^{-4} {\rm mpc}.
\label{eq:defrg}
\eeq

We define $\boldsymbol{\chi}$ to be the dimensionless spin angular momentum vector 
of the SBH,
\beq
\mathbf{J} = \boldsymbol{\chi}\left({G\mh^2\over c}\right),\ \ \ \ 0\le\chi\le 1.
\eeq
The standard (no-hair) relation between $J$ and the quadrupole moment
is
\beq
Q=-{1\over c} {J^2\over\mh}.
\label{eq:defq}
\eeq
We adopt this relation below unless otherwise noted.

In the regime of interest, stellar orbits around the SBH
can be approximated as Keplerian ellipses that experience gradual
changes in their orbital elements,
due both to the effects of relativity and to perturbations from other stars.
Here we summarize the relevant sources of evolution and their
associated time scales under this approximation.

\subsection{Relativistic precession}
\subsubsection{In-plane precession}

In the orbit-averaged approximation, massless test particles orbiting
a black hole experience advance of the orbital periapse by an angle
\footnote{Throughout, changes per orbit are denoted
by $\delta$ and changes over an arbitrary time interval by
$\Delta$.}
\beq
\delta\varpi = A_S - 2A_J\cos i - {1\over 2} A_Q (1 - 3\cos^2i)
\eeq
per orbit, where the subscripts $S,J,Q$ denote the
effects due to the black holes's mass (i.e. the Schwarzschild
part of the metric), spin and quadrupole moment (the Kerr part
of the metric) respectively and $i$ is the orbital inclination,
defined as the angle between the SBH spin vector and the stellar
orbital angular momentum vector.
To lowest post-Newtonian PN order,
\begin{subequations}
\begin{eqnarray}
A_S &=& {6\pi\over c^2} {G\mh\over (1-e^2)a} \nonumber \\ 
&\approx& 12.4' (1-e^2)^{-1} {\tilde a}^{-1}, \\ 
A_J &=& {4\pi \chi\over c^3}\left[{G\mh\over (1-e^2)a}\right]^{3/2} \nonumber \\ 
&\approx& 0.115' (1-e^2)^{-3/2}\chi {\tilde a}^{-3/2}, \label{eq:defaj}
\\
A_Q &=& {3\pi \chi^2\over c^4}\left[{G\mh\over (1-e^2)a}\right]^{2} \nonumber \\ 
&\approx& 1.19'\times 10^{-3} (1-e^2)^{-2}\chi^2 {\tilde a}^{-2}
\label{eq:defaq}
\end{eqnarray}
\end{subequations}
where $e$ is the orbital eccentricity.
Since the Schwarzschild contribution exceeds in amplitude the spin- 
and quadrupole contributions 
to the in-plane precession for $a (1-e^2)\gap 10^{-4}$ mpc $\approx r_g$,
advance of the periapse does not contain much useful information about the 
SBH spin \cite{Will-08}.

We define the precession time scale due to the Schwarzschild
term alone as
\begin{subequations}
\begin{eqnarray}
t_{S} &\equiv& \left[{A_S(a,e)\over \pi P(a)}\right]^{-1} \\
&=& {P\over 6} {c^2 a\over G\mh} (1-e^2)  \\
&\approx& 1.29\times 10^3 {\rm yr} \left(1-e^2\right) \tilde a^{5/2}.
\end{eqnarray}
\label{eq:defT}
\end{subequations}
The Schwarzschild contribution to the in-plane precession 
is large enough to potentially be detectable via a few years' monitoring
of identified stars at $\sim 10$ mpc separations from the SBH
\cite{Fragile-00,Rubilar-01,Zucker-06}.

\subsubsection{Precession of orbital planes}

The gravitational field of a Kerr black hole is not spherically symmetric.
The dominant non-spherically-symmetric effect on test-particle orbits
is the coupling between the spin of the black  hole and the orbital
angular momentum of the particle, known in the weak-field limit
as the Lense-Thirring effect \cite{LT-18,Wilkins-72}.
Again to lowest PN order, the change per orbit of the nodal angle $\Omega$ is
\cite{Will-08}
\beq
\delta\Omega = A_J - A_Q \cos i 
\label{eq:dOmega}
\eeq
\footnote{
Here and elsewhere, we ignore the distinction between the
total orbital angular momentum, 
and the Newtonian angular momentum, which exhibits a slight
additional ``wobble'' \cite{Kidder-95}.
It is easy to show that the amplitude of the wobble is
tiny given the parameters considered here.}.
Among relativistic effects, precession of orbital planes depends only 
on $\mathbf{J}$ and $Q$.
We note that the frame dragging contribution in Eq.~(\ref{eq:dOmega}), 
which dominates at most distances of interest here, is independent of 
orbital inclination while the quadrupole-induced precession is
inclination-dependent.
Both precessions leave the inclination with respect to the black hole
spin unchanged.

\begin{figure}
\includegraphics[width=8.cm]{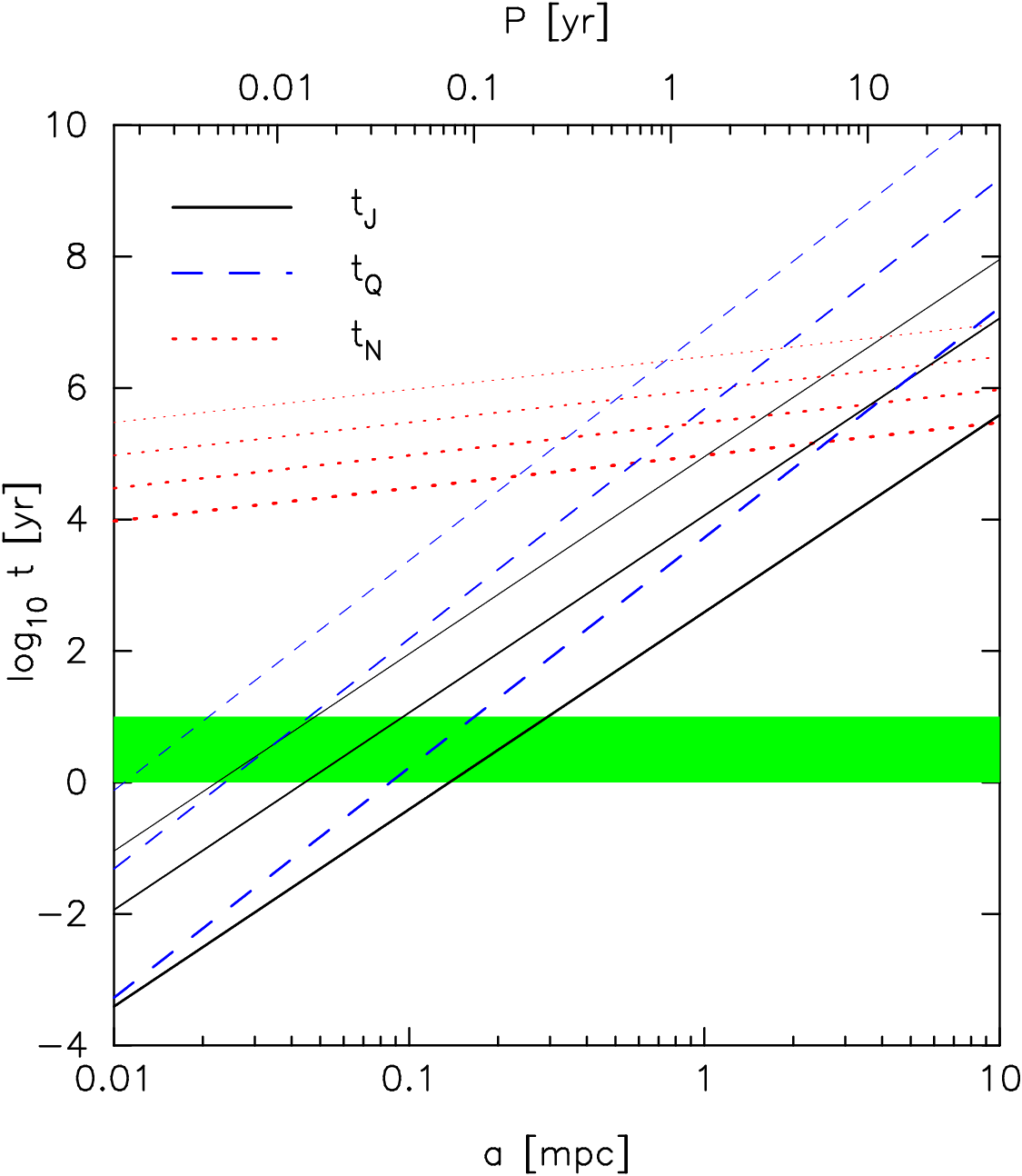}
\caption{\label{fig:at} 
Time scales associated with
precession of orbital planes about the Galactic supermassive black hole.
$t_J, t_Q$: precession time scales due to frame-dragging and to the quadrupole
torque from a maximally-spinning SBH. 
Line thickness denotes orbital eccentricity, from
$e=0.99$ (thickest) to $e=0.9$ and $e=0.5$ (thinnest).
$t_N$: approximate precessional time scale due to Newtonian perturbations 
from other stars, assumed to have one Solar mass.
Line thickness denotes total distributed
mass within $1$ mpc from the SBH, from $10^3\msun$
(thickest) to $1\msun$ (thinnest), assuming that density falls off
as $r^{-1}$.
Shaded (green) region shows range of interesting time intervals
for observation, $1\ {\rm yr}\le\Delta t\le 10$ yr.
}
\end{figure}

Defining the associated time scales as in Eq.~(\ref{eq:defT}), we find
\begin{subequations}
\begin{eqnarray}
t_J &=& {P\over 4\chi} \left[{c^2 a(1-e^2)\over G\mh}\right]^{3/2} \\
&\approx& 1.39\times 10^5 {\rm yr} \left(1-e^2\right)^{3/2}\chi^{-1}\tilde a^3,\\
t_Q &=& {P\over 3\chi^2} \left[{c^2 a(1-e^2)\over G\mh}\right]^2 \\
&\approx& 1.34\times 10^7 {\rm yr} \left(1-e^2\right)^2 \chi^{-2} {\tilde a}^{7/2}.
\end{eqnarray}
\end{subequations}
Figure~\ref{fig:at} plots $t_J$ and $t_Q$ 
as functions of $a$ and $e$.

\subsection{Stellar perturbations}

If there is a star cluster around the SBH, the smooth
contribution to the gravitational force from the distributed mass
breaks the degeneracy between radial and
angular periods in the classical Kepler problem, 
causing an in-plane precession, in the opposite sense to the
relativistic periastron advance.
Assuming that the stellar mass density follows
$r^{-\gamma}$, with $r$ the distance from the SBH,
the advance of orbital periapse in one period is
\beq
\delta\varpi \approx 2\pi {M_\star(a)\over\mh}\sqrt{1-e^2} F(\gamma)
\eeq
where $M_\star(r)$ is the distributed mass enclosed within radius $r$ and
$F=(3/2, 1)$ for $\gamma=(0,1)$ \cite{MV-10}.
Setting $F\approx 1$, the associated time scale is
\begin{subequations}
\begin{eqnarray}
t_M &\approx& {P\over 2} {\mh\over M_\star} (1-e^2)^{-1/2} \\
&\approx& 3.0\times 10^6 {\rm yr} \tilde M_\star^{-1} \tilde a^{\gamma-3/2}  
(1-e^2)^{-1/2}
\end{eqnarray}
\end{subequations}
where $\tilde M_\star$ is the stellar mass within 1 mpc
in units of the solar mass.
This time scale is long compared with the time $t_S$ for
relativistic periastron advance, Eq.~(\ref{eq:defT}),
at all radii of interest unless $M_\star$ is unphysically large.

The discrete nature of the stellar cluster adds an additional,
non-spherically-symmetric component to the gravitational
potential, which can induce precession in orbital planes
that mimics the effects of frame-dragging and quadrupole torques.
In the case that the time scale associated with this 
precession is long compared with both the radial period 
(Eq.~\ref{eq:period}) 
and with the time scale for in-plane periastron advance
(Eq.~\ref{eq:defT}) (assumptions that will be verified below), 
orbits around the SBH respond to the finite-$N$ component of the
gravitational force as if they were annuli, 
changing their orientations but not their eccentricities
(``vector resonant relaxation''; \cite{Rauch-96}).

Here we estimate the rate of precession due to finite-$N$ 
stellar perturbations, adopting a purely Newtonian model for 
star-star interactions.

Let $q\equiv \ms/\mh$ be the ratio between stellar mass
and SBH mass,
$N$ the number of stars and/or stellar remnants
in the region contained within a test star's orbit
(a more precise definition of $N$ is adopted in \S IV)
and $L_c$ the angular momentum of a circular orbit of the same energy as that
of the test star.
In the vector resonant relaxation (RR) regime,
orbital angular momenta evolve
approximately as \cite{Rauch-96}
\beq
{|\Delta\mathbf{L}|\over L_c} \approx \beta_v q\sqrt{N} 
{\Delta t\over P} 
\label{eq:vrr}
\eeq
for a time $\Delta t\lap t_{\rm coh}$, where $\beta_v$ is
a constant of order unity and $t_{\rm coh}$
is the time scale associated with the most rapid process
that randomizes orbital planes, thus breaking the coherence.

In the absence of GR effects, the only source of coherence-breaking
is the stellar perturbations themselves (``self-quenching''),
for which $t_{\rm coh}=t_N$, where $t_N$ is defined by the
condition $|\Delta\mathbf{L}|/L_c(t_N)=1$.
On time scales long compared to $t_N$, and in the absence
of frame-dragging or other torques,
orbital orientations would evolve approximately as
\beq
{|\Delta\mathbf{L}|\over L_c} \approx \beta_v q\sqrt{N} 
{\sqrt{t_{\rm coh}\Delta t}\over P} ,
\label{eq:coh}
\eeq
i.e. as $(\Delta t)^{1/2}$ rather than as $(\Delta t)^1$.
Fig.~\ref{fig:at} shows that $t_N$ is $\gap 10^4$ yr
for reasonable models of the stellar cluster, much longer
than the $\sim 10$ yr time scales of interest here; hence
the self-quenched regime is irrelevant in what follows.

However at some radius, $t_N$ will exceed the time scales 
associated with GR precession of orbital planes, and
the precession rate will be given by the expressions derived in
the previous section rather than by Eq.~(\ref{eq:vrr}).
To estimate this radius, 
we begin by expressing the GR precessional time scales defined above in terms
of the ``penetration parameter'' 
${\varrho}\equiv(1+e)r_{p}/r_{g}>1$, where $r_{p}$ is
the Keplerian orbital periapse distance and $r_g$ is 
defined in Eq.~(\ref{eq:defrg}). The results are:
\begin{subequations}
\begin{eqnarray}
t_J &=& {1\over 4} \varrho^{3/2}\chi^{-1}P, \\
t_Q &=& {1\over 3} \varrho^2\chi^{-2}P 
\end{eqnarray}
\label{eq:tJtQ}
\end{subequations}
and the vector RR time scale itself is
\beq
t_N \approx {1\over q\sqrt{N}} P; \tag{14c}
\label{eq:deftv1}
\eeq
the latter expression is true only up to ${\cal O}(1)$
factors which have to be derived from simulations. 

Since 
\beq
{t_{Q}\over t_{J}} = (4/3\chi)\varrho^{1/2}\ge4/3\,,
\eeq
precession due to frame-dragging is everywhere faster than 
precession due to the quadrupole torque.

The condition that frame dragging dominate
stellar perturbations is approximately
\begin{equation}
{t_N\over t_J} \approx {4\chi \over q\sqrt{N}\varrho^{3/2}} >1\,.
\end{equation}
Fig.~\ref{fig:at} shows that for $\chi=1$, this condition
is satisifed inside $\sim 1$ mpc for reasonable values of the enclosed mass.
For $\chi<1$, the critical radius is smaller.
This justifies looking at stellar perturbations as a source
of ``noise'' in tests of GR.

\begin{figure}
\includegraphics[width=8.5cm]{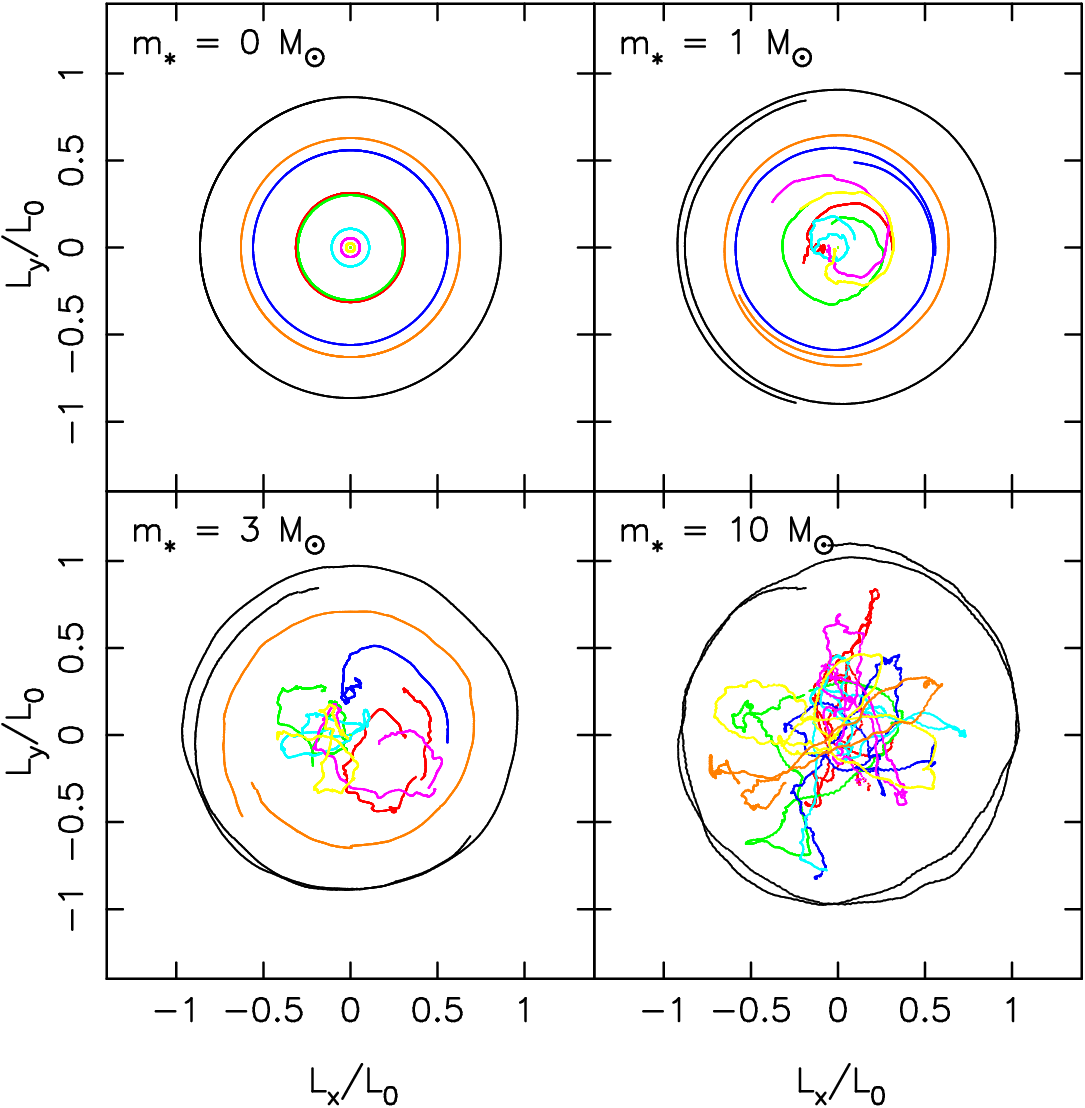}
\caption{\label{fig:four} Evolution of orbital planes
in a cluster of eight stars orbiting about the Galactic center SBH,
for an elapsed time of $2\times 10^6$ years.
The SBH rotates about the $z$-axis with maximal spin.
Four different values were assumed for the stellar masses $m_\star$, 
as indicated.
Stars were placed initially on orbits with semi-major axis 
$2$ mpc and eccentricity $0.5$ and with random orientations.
}
\end{figure}

\subsection{Comparing relativistic and Newtonian precessions}

Precession of orbital planes induced by stellar perturbations
differs qualitatively from precession due to frame dragging
since it does not respect the direction of the SBH spin axis.
Fig.~\ref{fig:four} shows the results of a set of $2\times 10^6$ yr
integrations (using the algorithm described in \S\ref{sec:nbody})
that illustrate the difference.
When the stellar masses are set to zero, orbital angular momenta
exhibit the uniform precession about the SBH's spin axis 
associated with frame dragging; 
when stellar masses are increased,
the orbital angular momentum vectors
move quasi-randomly about the unit sphere.

In comparing GR precession  with that due to
stellar perturbations, it is therefore useful to have a measure
of orientation that is invariant to the direction of
the SBH spin.
We adopt $\Delta\theta$, defined as the angle between
the initial and final orbital angular momentum vectors:
\beq
\cos\Delta\theta = {\mathbf{L}_i\cdot\mathbf{L}_f\over L_iL_f}.
\label{eq:defdtheta}
\eeq
On time scales of relevance here, $|\mathbf{L}|$ is conserved,
i.e. $L_i\approx L_f$, because the time scales for
both non-resonant and resonant relaxation are much longer
than $10$ yr.

Precession induced by GR changes only the nodal angle $\Omega$
(Eq.~\ref{eq:dOmega}).
Since 
\beq
\mathbf{L}=L\left(\sin\Omega\sin i\mathbf{e_x} + 
\cos\Omega\sin i \mathbf{e_y} + \cos i \mathbf{e_z}\right),
\eeq
where the $z$-axis is parallel to the SBH spin vector,
$L_i=L_f$ implies
\beq
\cos\Delta\theta_{GR} = \cos^2 i + \sin^2 i\cos\Delta\Omega
\eeq
which for small $\Delta\Omega$ is
\beq
\Delta\theta_{GR}\approx \sin i \Delta\Omega.
\label{eq:dtGR}
\eeq

In the case of star-star perturbations, we need to express
$\Delta\theta$ in terms of $\Delta\mathbf{L}/L_c$.
By definition,
\begin{equation}
\left|\Delta\mathbf{L}\right|^2=L_{i}^{2}+L_{f}^{2}
- 2L_{i}L_{f}\cos\Delta\theta
\end{equation}
so 
\begin{subequations}
\begin{eqnarray}
{\left|\Delta\mathbf{L}\right|^2\over L_{c}^2}&=& 
2{L^2\over L_c^2} \left(1-\cos\Delta\theta\right) \\
&\approx& {L^2\over L_c^2} \left(\Delta\theta\right)^2 
\end{eqnarray}
\label{eq:vrr2}
\end{subequations}
where the last expression again assumes small $\Delta\theta$.

Specializing Eq.~(\ref{eq:dtGR}) to the case of frame dragging, 
we note that $\Delta\Omega$ is independent of $\cos i$.
Considering orbits with a single eccentricity $e$ and with
an isotropic distribution of inclinations,
the rms values of the angles in
Eq.~(\ref{eq:dtGR}) are therefore related by
\beq
\Delta\theta_J \approx \sqrt{2\over 3}\Delta\Omega
\eeq
or (cf. Eq.~\ref{eq:defaj})
\beq
\Delta\theta_J \approx 4\pi\sqrt{2\over 3}\chi \varrho^{-3/2} {\Delta t\over P}.
\label{eq:dtfd2}
\eeq

In the case of quadrupole-induced precession, $\Delta\Omega\propto\cos i$.
Again computing the rms values assuming random orientations gives
\beq
\Delta\theta_Q \approx \sqrt{6\over 5}\pi\chi^2 \varrho^{-2} {\Delta t\over P}.
\label{eq:dtq2}
\eeq

Finally, for stellar perturbations, we ignore a possible dependence
of $|\Delta\theta_N|$ on orbital eccentricity.
In an isotropic cluster, the orbital angular momenta at any 
energy ($\sim$ radius) are distributed as 
$n(L)\mathrm{d}L=2L\mathrm{d}L/L_c^2$, so
$\left\langle L^{2}\right\rangle /L_{c}^{2}=1/2$,
and the rms values in Eq.~(\ref{eq:vrr2})
are related by
\beq
\Delta\theta_{N} \approx \sqrt{2}{|\Delta\mathbf{L}|\over L_{c}}.
\label{eq:dtvrr}
\eeq
Using Eq.~(\ref{eq:vrr}), this can be written
\beq
\Delta\theta_N \approx \sqrt{2}\beta_v q\sqrt{N} {\Delta t\over P}\, .
\label{eq:dtvrr1}
\eeq
Eilon et al. \cite{Eilon-09} give $\beta_v\approx 1.8$
as an average value for a cluster with isotropically distributed
velocities, if $N$ is defined as the number of stars within
a sphere of radius $r=a$.
We can then write a slightly more accurate
definition of the vector RR time scale
(again defined as the time such that $\Delta\theta_N=\pi$),
\beq
t_N = {\pi\over\sqrt{2}\beta_v} {1\over q\sqrt{N}}P 
\approx {1.2\over q\sqrt{N}}P.
\label{eq:deftv2}
\eeq
This is the expression plotted in Fig.~1.

We have assumed that precession is due either to GR spin effects
or to stellar perturbations. In reality, one expects vector RR
to be quenched somewhat by coherence-breaking due to GR precession
even at radii where $t_N<t_{J,Q}$.

Equating (\ref{eq:dtfd2}) with 
(\ref{eq:dtvrr1}),
we obtain an approximate expression for the radius at which
frame-dragging dominates stellar perturbations:
\beq
\varrho^{3/2}\sqrt{N} \approx {4\pi\over\sqrt{3}\beta_v}{\chi\over q}
\label{eq:rcrit}
\eeq
i.e.
\beq
r_{\rm crit} \approx 1 \text{mpc} (1-e^2)^{-1} \chi^{2/3}
\left({N_{\rm crit}\over 30}\right)^{-1/3}
\left({m_\star\over 10\msun}\right)^{-2/3}
\eeq
where $N_{\rm crit}$ is the number of stellar perturbers
within $r_{\rm crit}$ of mass $m_\star$ each.
We evaluate this expression in \S\ref{sec:models}
after specifying a model for the stellar distribution,
and in \S\ref{sec:results} we present the results of 
full $N$-body simulations that allow more precise estimates
of $r_{\rm crit}$.

\bigskip
\section{\label{sec:nbody}$N$-body treatment}

Integrations of the $N$-body equations of motion were carried out using
algorithmic regularization \cite{MT-99a,MT-99b} implemented with a chain
structure \cite{MA-93} and the time-transformed leapfrog \cite{MA-02}.
The algorithm produces exact trajectories for Newtonian two-body 
motion and regular results for close encounters involving arbitrary
numbers of bodies.
Velocity-dependent forces were included via a generalized mid-point
method \cite{MM-06}; the {\tt ARCHAIN} code \cite{MM-08} also incorporates 
pairwise post-Newtonian forces for non-spinning particles of orders 
up to and including PN2.5 \cite{Soffel-89}.
We included PN terms in the interactions between the SBH particle 
and the $N-1$ ``star'' particles.
All $N$ particles were included at all times in the chain.
Accumulated energy errors were never more than a few parts in $10^{10}$.

We modified {\tt ARCHAIN} to include the lowest-order contributions of the 
SBH's spin and quadrupole moment to the motions of the stars.
In the covariant spin supplementary condition (SSC) gauge
\cite{Kidder-95}, the spin-related, $N$-body accelerations $\mathbf{a}_J$ 
are
\begin{widetext}
\begin{subequations}
\begin{eqnarray}
\mathbf{a}_{J,1} &=& -{3G^2\mh\over c^3} \sum_{j\ne 1} {m_j\over r_{1j}^3}
\bigg\{\left[\mathbf{v}_{1j}-\left(\mathbf{n}_{1j}\cdot\mathbf{v}_{1j}\right)
\mathbf{n}_{1j}\right]
\times\boldsymbol{\chi} - 
2\mathbf{n}_{1j}\left(\mathbf{n}_{1j}\times
\mathbf{v}_{1j}\right)\cdot\boldsymbol{\chi}
\bigg\},  \label{eq:a1} \\
\mathbf{a}_{J,j} &=& {2G^2\mh^2\over c^3r_{1j}^3} 
\bigg\{\left[2\mathbf{v}_{1j}-3\left(\mathbf{n}_{1j}\cdot\mathbf{v}_{1j}\right)
\mathbf{n}_{1j}\right] \times\boldsymbol{\chi} - 
3\mathbf{n}_{1j}\left(\mathbf{n}_{1j}\times
\mathbf{v}_{1j}\right)\cdot\boldsymbol{\chi}
\bigg\}, \label{eq:aj} \\
\dot{\boldsymbol{\chi}} &=& {G\over 2c^2} \sum_{j\ne i}{m_j\over r_{ij}^2} 
\left[\mathbf{n}_{1j}\times\left(3\mathbf{v}_1-4\mathbf{v}_j\right)\right]
\times\boldsymbol{\chi}, \label{eq:alphadot} \\
r_{ij}&\equiv& \left|\mathbf{x}_i - \mathbf{x}_j\right|,\ \ \ 
\mathbf{x}_{ij}\equiv \mathbf{x}_i-\mathbf{x}_j,\ \ \ 
\mathbf{n}_{ij} = \mathbf{x}_{ij}/r_{ij}, \ \ \ 
\mathbf{v}_{ij} \equiv \mathbf{v}_i - \mathbf{v}_j.
\end{eqnarray}
\end{subequations}
\end{widetext}
Here, particle number 1 is the SBH and particles $j$, $2\le j\le N$
are the stars.
The ``linear momentum'' that is conserved by these equations is
\begin{equation}
\mathbf{P} = \sum_i m_i\mathbf{v}_i +
{G\over c^2}\sum_{ij} {m_i \over 2r_{ij}^3}
\left(\mathbf{x}_{ij}\times {\mathbf{J}}_j\right).
\end{equation}

Adopting  Eq.~(\ref{eq:defq}) 
for the SBH quadrupole  moment,
the equation of motion for the $j$th particle has the additional
term $\mathbf{a}_{Q,j}$, where
\begin{widetext}
\begin{equation}
\mathbf{a}_{Q,j} = +{3\over 2}\chi^2{G^3\over c^4} {\mh^3\over r^4}
\left[5\mathbf{n}_{1j}
\left(\mathbf{n}_{1j}\cdot\hat{\mathbf{J}}\right)^2 
- 2\left(\mathbf{n}_{1j}\cdot\hat{\mathbf{J}}\right)\hat{\mathbf{J}} 
- \mathbf{n}_{1j}\right], \ \ \ \ 
\hat{\mathbf{J}} \equiv \mathbf{J}/J.
\end{equation}
\end{widetext}

\section{\label{sec:models}Models for the stellar distribution}
\subsection{Observational constraints}

The distribution of stars and stellar remnants at distances
$\ll 1$ pc from the Galactic center SBH is poorly understood.
Only the brightest stars in the inner parsec have been identified, 
via speckle or adaptive optics imaging and spectroscopy in the 
near-infrared bands \cite{Eckart-93,Genzel-96}.
Most of these stars appear to belong to one of two distinct populations:
(1) ``early-type'' (ET) stars -- apparently normal, upper-main-sequence 
giant stars
of O and B spectral types with inferred masses of $7-80\msun$ and ages less 
than the main-sequence turnoff age, i.e. $O[10^1-10^2]$ Myr; and
(2) ``late-type'' (LT) stars -- old, metal-rich, M, K and G-spectral-type 
giant (post-main-sequence) stars with ages $O[10^0-10^1]$ Gyr and masses 
$1-2\msun$.
The density of ET stars increases steeply toward the SBH and these stars
account for a large part of the total luminosity of the central cluster,
but their total numbers are small, roughly $10^2$ in the inner
$0.1$ pc \cite{Paumard-06,Gillessen-09a,Buchholz-09} with few if any
on orbits that bring them within $\sim 10$ mpc from the SBH,
making them unlikely candidates either as test stars for observing GR 
spin effects or as perturbers of the test stars.

The LT stars on the other hand are believed to be characteristic 
of the dominant, old population;
roughly 6000 LT stars have been identified in the inner $\sim 0.5$ pc 
and their K-band luminosity function suggests a roughly continuous star
formation history over the  last $\sim 10$ Gyr \cite{Genzel-03,Buchholz-09}.
In spite of their large numbers, the observed LT stars
appear to be weakly concentrated toward the SBH.
Number counts complete to $K\approx 15.5$ (corresponding to the sub-giant
phase for $1\msun$ stars)
reveal a projected density that is flat or declining
inside a projected distance of $\sim 0.5$ pc from the SBH 
\cite{Buchholz-09,Do-09,Bartko-09}.
While the existence of four LT stars on very tight 
($5\ {\rm mpc}\lap a\lap 20\ {\rm mpc}$)
orbits around the SBH has been established 
\cite{Gillessen-09a},
deprojection of the binned surface density profile implies a central space
density that is consistent with zero at distances smaller than
$\sim 0.1$ pc from the SBH \cite{Merritt-09}.

The low density of LT stars in the inner parsec is not well understood.
If the time scale for exchange of orbital kinetic energy 
between stars (the two-body relaxation time; \cite{Spitzer-86}) 
is shorter than several Gyr, 
one expects the stellar distribution to have attained a quasi-steady-state
distribution of the form
$n(r)\sim r^{-\gamma}$, $3/2\lap\gamma\lap 7/4$ \cite{BW-76,BW-77}
within the SBH gravitational influence radius,
$r_{\rm infl}\equiv G\mh/\sigma_\star^2\approx 10^0$ pc.
This is clearly not observed \cite{Do-09}, 
suggesting either that the relaxation
time exceeds $\sim 10$ Gyr throughout the inner parsec, 
or that the brightest stars have been hidden from view or destroyed.
Collisions with main-sequence stars or stellar remnants
can remove the outer envelopes of red-giant stars, 
potentially explaining the low observed density of giants
\cite{Genzel-96}.
However this mechanism only appears to be effective
at distances less than $\sim 0.1$ pc from the SBH
\cite{Alexander-99,Freitag-06,Dale-09}, even assuming a high
density for the colliding populations (an assumption for which there is 
currently no observational support).
Even at these radii, collisions would seem to be ineffective
at explaining the depletion of stars down to magnitudes as faint as 15.5
\cite{Dale-09}.

It has been argued that the stellar initial mass function (IMF) may
have been strongly truncated below $\sim 3\msun$ in the Galactic 
center region \cite{Naya-05,Maness-07}.
These are just the stars that would dominate the K-band number
counts now \cite{Dale-09}.

At the high-mass end, standard IMFs \cite{Scalo-86,Kroupa-01}
predict that $\sim 0.1\%$ of
stars have initial masses greater than $20\msun$, ending their
short lives as $\sim 5-15\msun$ black holes (BHs).
The BHs are expected to segregate nearer to the SBH
than the lower-mass components 
(stars, white dwarves, neutron stars \cite{Morris-93}), 
possibly dominating the total
number density inside $\sim 1$ mpc \cite{HA-06,Freitag-06,AH-09}
and providing the bulk of the perturbations acting on the observed
stars in this region.
However if the observed distribution of late-type stars is a guide,
the two-body relaxation time may be too long 
for establishment of a mass-segregated distribution
 \cite{Freitag-06,MS-06,Merritt-09}.
If this is the case, there is no compelling reason
to assume that the ratio of BHs to stars is as large as implied
by the mass-segregated models.

Proper motion studies of large samples of LT stars in the 
inner parsec \cite{Eckart-97,Schoedel-09} yield dynamical 
constraints on the 
distributed mass (stars, stellar remnants, gas etc.) in this region.
The proper motion data robustly require an extended mass of 
$\sim 0.5-1.5\times 10^6\msun$ within the central parsec
\cite{Schoedel-09}.
However these data do not strongly constrain the radial dependence
of the distributed mass density nor the amount of mass
on the mpc scales of interest here.

\subsection{Parametrized models}

Given these uncertainties, we explored a range of different
models for the distribution of stars and stellar remnants near
the Galactic SBH.
We define $M_\star$ as the distributed mass
within 1 mpc from the SBH and $\tilde M_\star \equiv M_\star/\msun$.
We idealize the stellar populations in this region as consisting 
of just two components: $1\msun$ main-sequence (MS) stars 
and $10\msun$ BHs.
The first population is assumed to be amenable to astrometric
monitoring, and all discussions of orbital evolution presented below 
will refer to this population.
While the orbits of the BHs are also allowed to evolve in our models,
we do not describe that evolution in what follows.

In addition to $M_\star$, three additional parameters define
the initial distributions of stars and stellar remnants
in our models:

\begin{itemize}
\item the power-law index $\gamma$ describing the number density profiles,
$n(r)\propto r^{-\gamma}$;
$\gamma$ is assumed to be the same for both MS stars and BHs;
\item the (number) ratio $\cal R$ of BHs to MS stars, i.e.
${\cal R} = N_{\rm BH}/N_{\rm MS}$;
\item the velocity anisotropy $\beta$, defined such that 
$\sigma_r^2/\sigma_t^2=(1-\beta)^{-1}$, where $\sigma_r$
and $\sigma_t$ are respectively the 1d velocity dispersions
in directions parallel and perpendicular to the radius vector;
$\beta=0$ corresponds to isotropy.
\end{itemize}

\begin{table}
\begin{center}
\caption{ \label{tbl-1}Parameters of the $N$-body models}
\begin{ruledtabular}
\begin{tabular}{rrrrrrrr}
$\gamma$ & $\beta$ & $\tilde M_\star$ & $\cal R$ & ${\tilde a}_{\rm max}$ &
$N_{\rm MS}$ & $N_{\rm BH}$ & $N_{\rm rand}$ \\
\tableline

0 & -1  & 10 & 0   & 4.0 & 159 & 0  & 6   \\
0 & -1  & 10 & 0.1 & 4.0 & 79  & 8  & 12  \\
0 & -1  & 10 & 1   & 4.0 & 14  & 15 & 70  \\

1 & -1  & 10 & 0   & 4.0 & 119 & 0  & 8   \\
1 & -1  & 10 & 0.1 & 4.0 & 59  & 6  & 16  \\
1 & -1  & 10 & 1   & 4.0 & 10  & 11 & 90  \\

1 & 0   & 30 & 0   & 3.5 & 183 & 0  & 6   \\
1 & 0   & 30 & 0.1 & 4.0 & 119 & 12 & 8   \\
1 & 0   & 30 & 1   & 4.0 & 21  & 22 & 45  \\

2 & -1  & 30 & 0   & 4.0 & 119 & 0  & 8   \\
2 & -1  & 30 & 0.1 & 4.0 & 59  & 6  & 15  \\
2 & -1  & 30 & 1   & 4.0 & 10  & 11 & 100 \\

2 & 0   & 10  & 1   & 4.0 & 4  & 3  & 300 \\

2 & 0   & 30  & 0   & 4.0 & 119 & 0  & 8   \\
2 & 0   & 30  & 0.1 & 4.0 & 59  & 6  & 15  \\
2 & 0   & 30  & 1   & 4.0 & 10  & 11 & 100 \\

2 & 0   & 100 & 0   & 1.75& 174 & 0  & 6   \\
2 & 0   & 100 & 0.1 & 4.0 & 179 & 19 & 6   \\
2 & 0   & 100 & 1   & 4.0 & 36  & 36 & 30  \\

2 & 0.5 & 100 & 0   & 1.75& 174 & 0  & 6   \\
2 & 0.5 & 100 & 0.1 & 4.0 & 179 & 19 & 6   \\
2 & 0.5 & 100 & 1   & 4.0 & 36  & 36 & 30  \\
\end{tabular}
\end{ruledtabular}
\end{center}
\end{table}

For $M_\star$ we adopted one of the three values $(10,30,100)\msun$;
the latter value is roughly the enclosed 
mass predicted by the relaxed, mass-segregated models cited above.
For $\gamma$ we considered the values $(0,1,2)$; $\gamma=0$
corresponds to a constant density in the inner mpc, roughly
what is observed in the projected density of LT stars,\footnote{
A flat, $\gamma=0$ distribution all the way out to $\sim 1$ pc
is inconsistent with the observations. For the $\gamma=0$ models,
we are assuming (as in the other models) that the flat distribution 
is local to a volume $\ll 1$ pc, and normalizing the total mass
to $(10,30,100)\msun$ inside $1$ mpc.}
 while
$\gamma=2$ is approximately the value expected for a mass-segregated
population around a SBH.
Adopting the proper-motion result that the distributed mass
within $1$ pc is $\sim 10^6\msun$ \cite{Schoedel-09},
the implied mass inside 1 mpc is $\sim 10^0(10^3)\msun$
for $\gamma = (1,2)$.
We therefore associated larger values
of $\tilde M_\star$ with larger values of $\gamma$, although
as noted above, the proper-motion data do not directly constrain
the mass distribution on mpc scales.
${\cal R}$ was set to $(0,0.1,1)$; ${\cal R}=1$ is roughly the
largest value predicted in the mass-segregated models assuming
a standard IMF, while
${\cal R}\approx 10^{-3}$ is expected in the absence of any 
mass segregation.
For the small ($\sim 10^2$) total particle numbers in the $N$-body
simulations, ${\cal R}=0$ is essentially the same as ${\cal R}=10^{-3}$.

\begin{figure}
\includegraphics[width=8.cm]{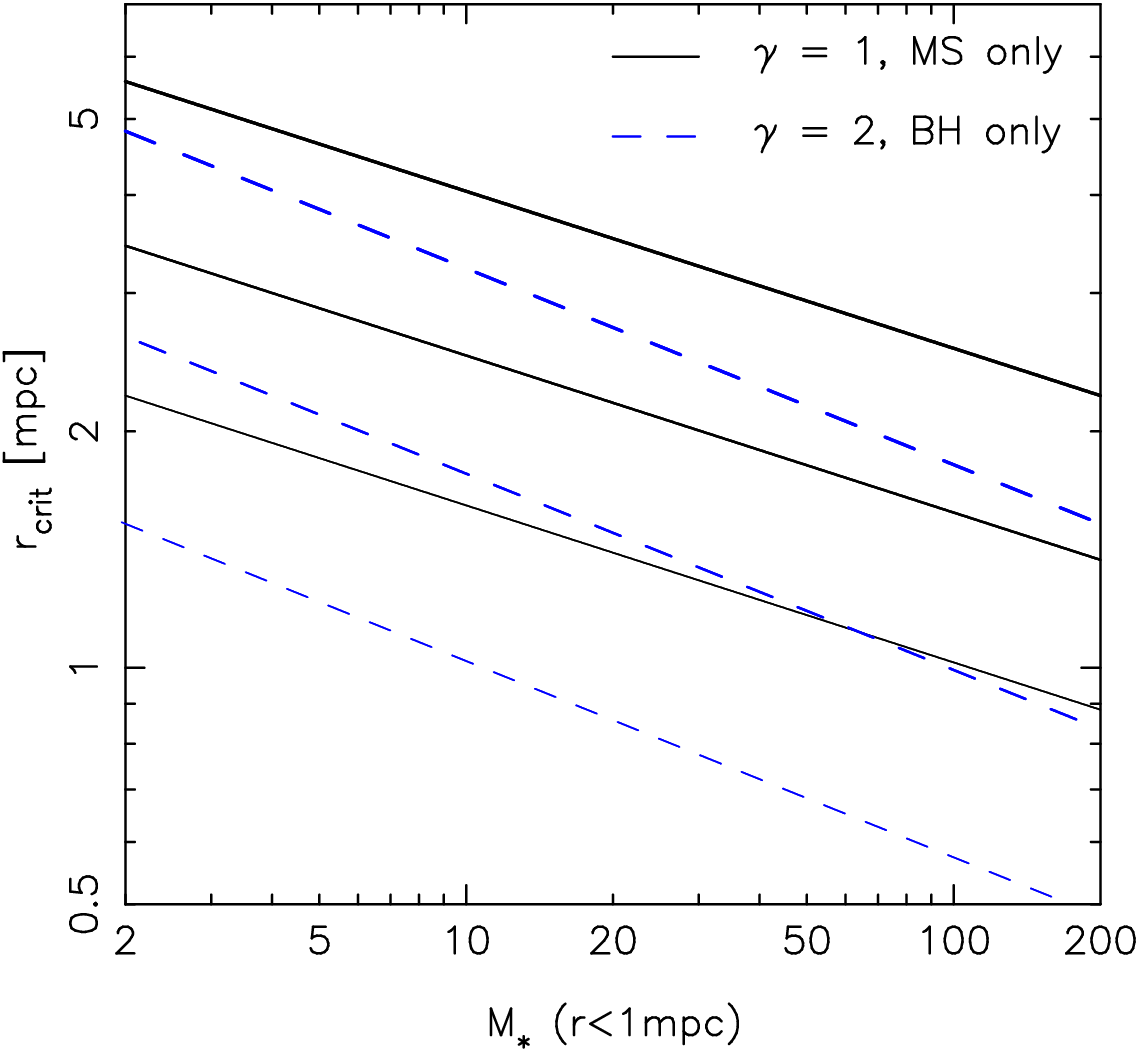}
\caption{\label{fig:rcrit} 
Approximate value of the radius at which stellar perturbations
match frame dragging in terms of their ability to change the direction
of orbital angular momenta (Eq.~\ref{eq:rcrit}).
Two models for the stellar cluster are shown:
a steeply-rising density profile, $\gamma=2$, that is dominated by 
$10\msun$ stellar black holes; and a shallower density profile,
$\gamma=1$, dominated by $1\msun$ main-sequence stars.
Horizontal axis is the total distributed mass within $1$ mpc from the SBH.
Line widths denote SBH spin: $\chi=1$ (thickest), $\chi=0.3$,
and $\chi=0.1$ (thinnest).
}
\end{figure}

A steady-state orbital distribution in a point-mass potential
requires $\beta<\gamma-1/2$, i.e., isotropic velocity distributions
are not permitted when $\gamma<0.5$: the distribution of orbital
eccentricities must be biased toward small values when the
spatial distribution is flat.
The Galactic center proper motion data cited above \cite{Schoedel-09} 
suggest approximate isotropy in the (projected) inner parsec.
Theoretically, two-body encounters should drive the distribution
toward isotropy near the SMBh while at the same time populating 
the low-angular-momentum orbits, producing an isotropic density cusp. 
However since the cusp is not observed, 
it is not clear that relaxation has had sufficient time to
reduce anistropies to low values \cite{Merritt-09}.
We therefore considered non-zero values
of $\beta$ even when setting $\gamma=1$ or $2$.

The following distribution of orbital elements:
\begin{subequations}
\begin{eqnarray}
N(a,e^2) da de^2 &=& N_0 g(a) h(e^2) da de^2, \\
g(a) &=& a^{2-\gamma}, \\
h(e^2) &=& \left(1-e^2\right)^{-\beta}, \ \ \beta\le \gamma-1/2 
\end{eqnarray}
\end{subequations}
generates steady-state phase-space distributions with the
properties defined above.
Monte-Carlo realizations of the stellar positions and velocities
were generated from this expression given the parameters
$(\gamma,\beta,\tilde M_\star,{\cal R})$.
Relativistic corrections were ignored when generating the initial conditions.
We assumed that $N(a,e^2) = 0$ for $a>a_{\rm max}$;
in most of the simulations, $a_{\rm max}=4$ mpc, but smaller
values were adopted as necessary to limit the total number
of particles to $\sim 180$, since for larger $N$ the {\tt ARCHAIN}
routine was found to run very slowly.
Table~\ref{tbl-1} gives the composition of all the models
discussed below.

Stellar orbits were excluded from the initial conditions 
if their periapse fell below
$20 G\mh/c^2$, roughly the distance at which a solar-mass
star on the main sequence would be tidally disrupted.

For each set of parameters defining the initial distribution,
a set of different Monte-Carlo realizations was generated
and independently integrated forward in time.
The number $N_{\rm rand}$ of independent realizations was chosen
such that the total number of MS stars in the combined
set of integrations was $\gap 10^3$. 
In addition, some integrations were repeated with the masses of the MS
stars and BHs set to zero (leaving just the relativistic terms capable
of inducing evolution of the orbital planes); 
and with the SBH spin set to zero (leaving just the Newtonian perturbations).

Given these models for the stellar cluster, 
we can use Eq.~(\ref{eq:rcrit}) to compute the approximate
radii $r_{\rm crit}$ where frame-dragging begins to dominate
stellar perturbations.
Assuming that the density is dominated either by $1\msun$ MS
stars (${\cal R}\ll 1$) or by $10\msun$ BHs (${\cal R}\gap 1$),
we find:
\beq
\tilde r_{\rm crit} \approx
\left[{(43,14)\chi\over(1-e^2)^{3/2} 
\tilde M_{\star}^{1/2}}\right]^{2\over 6-\gamma} \text{mpc}
\eeq
where the first number in
parentheses refers to the MS cluster and the second to the BH cluster.
Fig.~\ref{fig:rcrit} plots $r_{\rm crit}$ vs. $\tilde M_\star$ for
various values of $\chi$ assuming $e=2/3$, the mean eccentricity
in an isotropic distribution.

\begin{figure*}
\includegraphics[width=8.cm,angle=-90.]{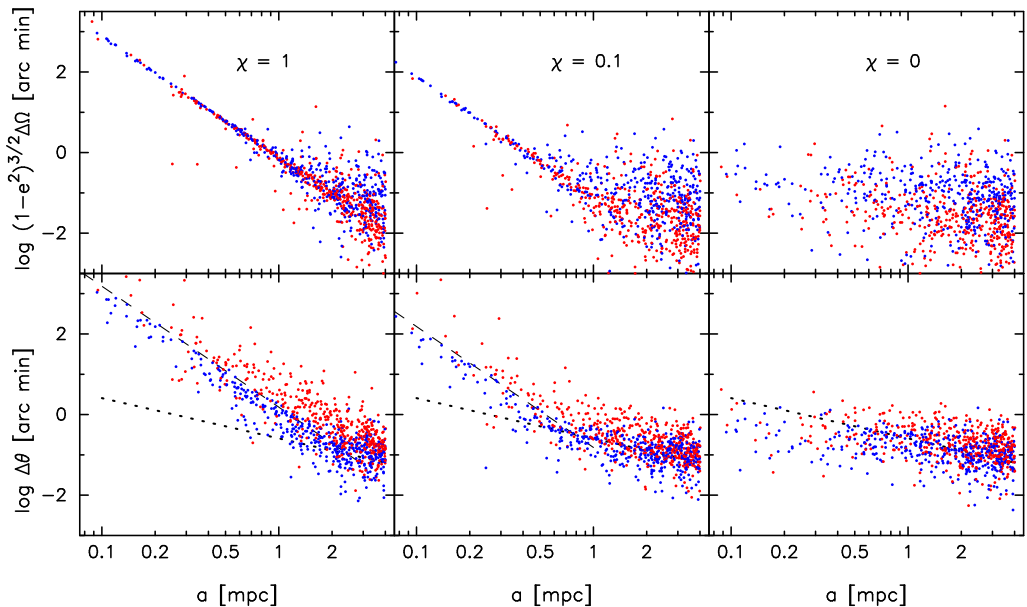}
\caption{\label{fig:M030} Changes over 10 years in the orientations
of stellar orbital planes, as measured via $\Delta\Omega$ (top)
and $\Delta\theta$ (bottom).
Parameters of the $N$-body models were $\gamma=2$, $\beta=0$, 
${\cal R}=1$, and $M_\star=30\msun$ ($N_{\rm MS}=10$, $N_{\rm BH}=11$).
Each point corresponds to a single star in a single integration;
red points are orbits with initial eccentricities $0.7< e\le 1$
and blue points have $0\le e\le 0.7$.
In the lower panels, dashed lines show Eq.~(\ref{eq:dtfd2}),
the frame-dragging precession, for $e=2/3$, 
and dotted lines show Eq.~(\ref{eq:dtvrr1}), the approximate
model for precession due to stellar perturbations.}
\end{figure*}

\begin{figure*}
\includegraphics[width=8.cm,angle=-90.]{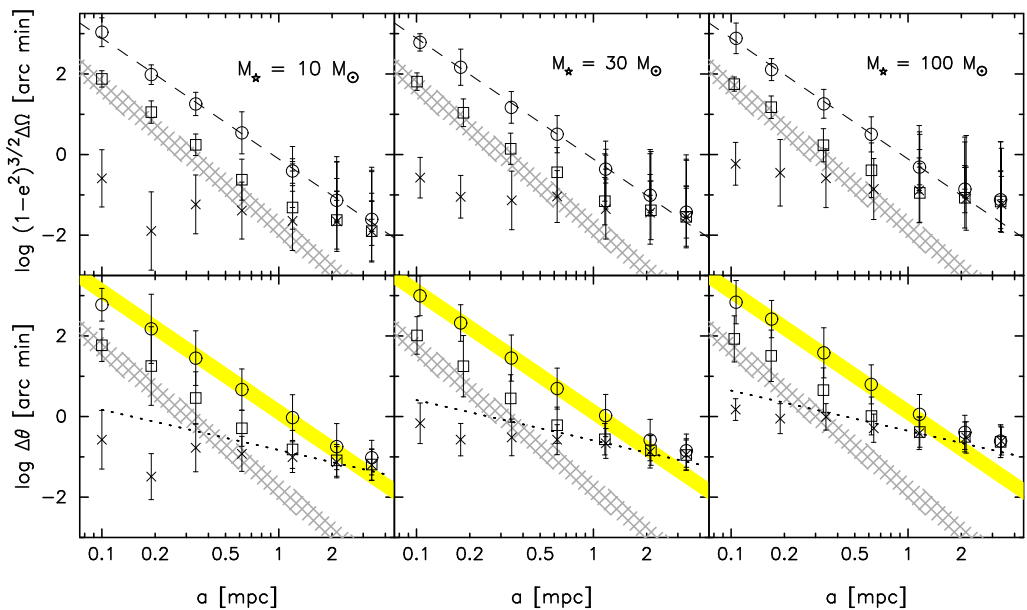}
\caption{\label{fig:bin} Similar to Fig.~\ref{fig:M030}
except that average values have been computed in bins of semi-major axis.
Three different $N$-body models are shown, differing 
in the distributed mass: $M_\star=(10,30,100)\msun$.
All models have $\gamma=2$, $\beta=0$, ${\cal R}=1$ as in Fig.~\ref{fig:M030}.
Open circles: $\chi=1$;
squares: $\chi=0.1$; 
crosses: $\chi=0$.
The predicted angular changes for $\chi=1$ due to frame-dragging 
are shown as the dashed lines in the upper histograms and as the yellow band
in the lower histograms.
The cross-hatched regions indicate the range of precession amplitudes
expected from the SBH quadrupole moment alone.
Dotted lines in the lower frames are 
the approximate model for stellar perturbations,
Eq.~(\ref{eq:dtvrr1}).}
\end{figure*}

\begin{figure*}
\includegraphics[width=15.cm]{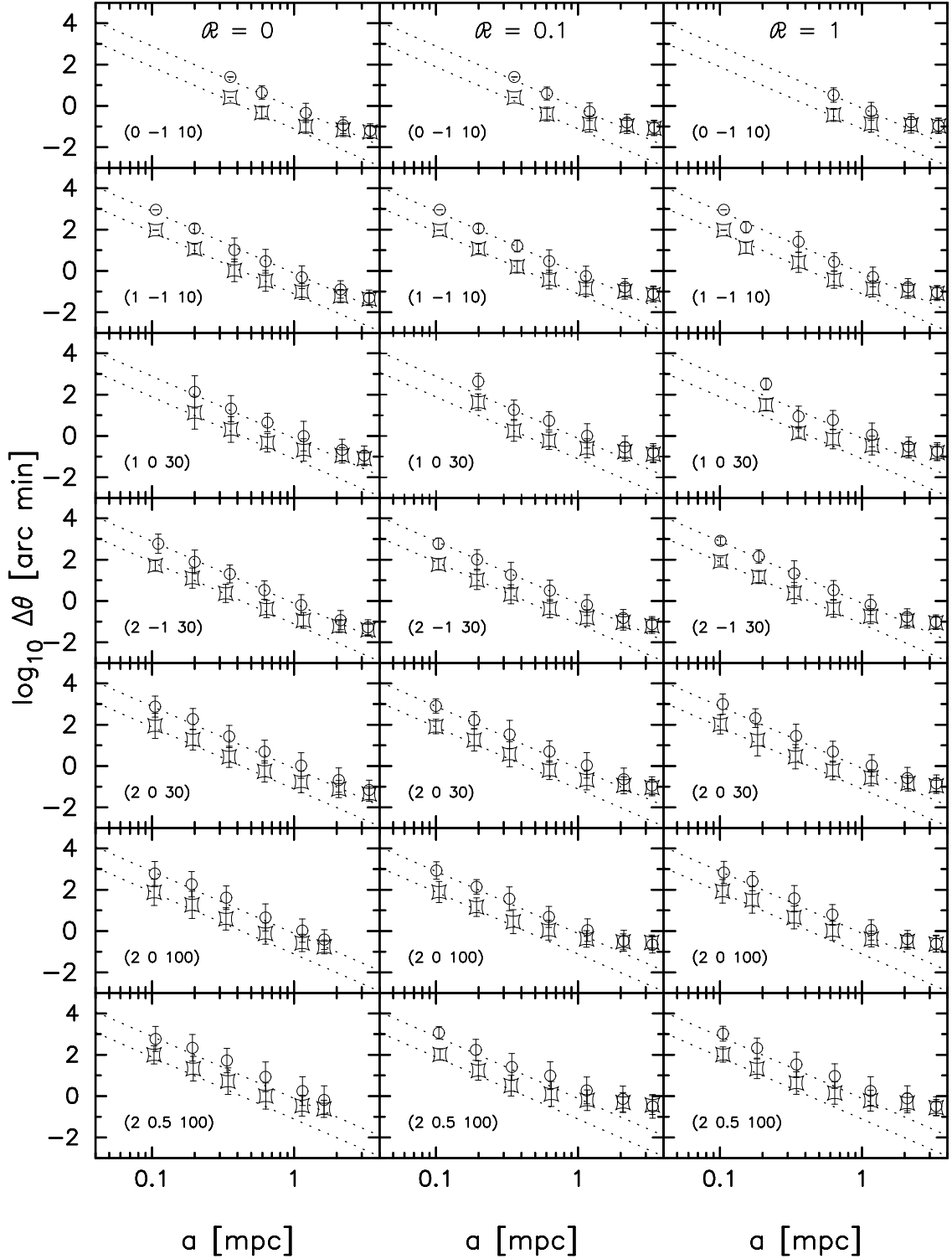}
\caption{\label{fig:all_bin} 
Changes over 10 years in the orientation
of stellar orbital angular momentum vectors, in $N$-body integrations 
of various models (Table~1).
The three columns correspond to three values ${\cal R}=(0,0.1,1)$
of the ratio of BHs to MS stars; the values of $\gamma, \beta$ and 
$\tilde M_\star$ that characterize the stellar distribution are given in the
lower left of each panel.
Circles are for integrations with $\chi=1$ and squares are for $\chi=0.1$.
Dashed lines show the expected, rms contribution to $\Delta\theta$
from frame dragging for both values of $\chi$.}
\end{figure*}

\section{\label{sec:results}Results}

In the $N$-body simulations, 
we characterized changes in stellar orbital orientations
in two ways: via $\Delta\Omega$, the change in the nodal angle
(defined with respect to the SBH equatorial plane);
and via the coordinate-independent quantity $\Delta\theta$,  
the angle between the initial and final
orbital angular momentum vectors (Eq.~\ref{eq:defdtheta}).
The nodal angle advances uniformly in time in response
to GR effects (Eq.~\ref{eq:dOmega}); 
furthermore for a given $\chi$ and $\Delta t$, 
the quantity $\left(1-e^2\right)^{3/2}\Delta\Omega$ depends
only on $a$ in the frame-dragging regime (Eq.~\ref{eq:defaj}, \ref{eq:dOmega}).

Fig.~\ref{fig:M030} plots $\left(1-e^2\right)^{3/2}\Delta\Omega$ and
$\Delta\theta$ vs. $a$ for each of the MS stars in a set of 
10-year integrations of models with $\gamma=2$, $\beta=0$, ${\cal R}=1$ and 
$\tilde M_\star=30$ and three different values of the SBH spin,
$\chi=(1,0.1,0)$.
Also shown are the predicted, rms values of $\Delta\theta$ from 
Eqs.~(\ref{eq:dtfd2}) (frame dragging) and~(\ref{eq:dtvrr1})
(stellar perturbations).
In this model cluster, in which stellar BHs dominate the total mass,
stellar perturbations dominate changes 
due to frame-dragging beyond
radii of $\sim 1 (0.3)$ mpc for $\chi=1(0.1)$.
As measured via $\Delta\theta$, GR effects are strongest for eccentric 
orbits, as expected, while the amplitude of the stellar perturbations
is not noticeably $e$-dependent.
The stellar perturbation model derived above is reasonably
good at predicting the mean value of $\Delta\theta$ in the integration
with $\chi=0$, although the observed dependence on $a$ appears
to be shallower than predicted for $a\lap 0.5$ mpc.

Because of the large scatter in the amplitude of stellar perturbations
at each $a$, the radius at which the GR signal clearly stands out
from the ``noise'' is somewhat smaller than would be predicted
from the rms values alone (Fig.~\ref{fig:rcrit}).

The dependence of these results on the amount of distributed mass
is shown in Fig.~\ref{fig:bin}, which summarizes results from integrations 
of models with $M_\star=(10,30,100)\msun$; other parameters are as in
Fig.~\ref{fig:M030}.
As $M_\star$ is increased, the amplitude of the ``noise''
from star-star perturbations increases, roughly in proportion
to $M_\star$.
These plots also indicate the expected amplitude of the
quadrupole-induced precession.
For $\chi=1$, stellar perturbations dominate changes due to the
quadrupole at radii beyond $\sim 0.5(0.3)$ mpc for
$M_\star=10(100)\msun$.

\begin{figure}
\includegraphics[width=8.5cm]{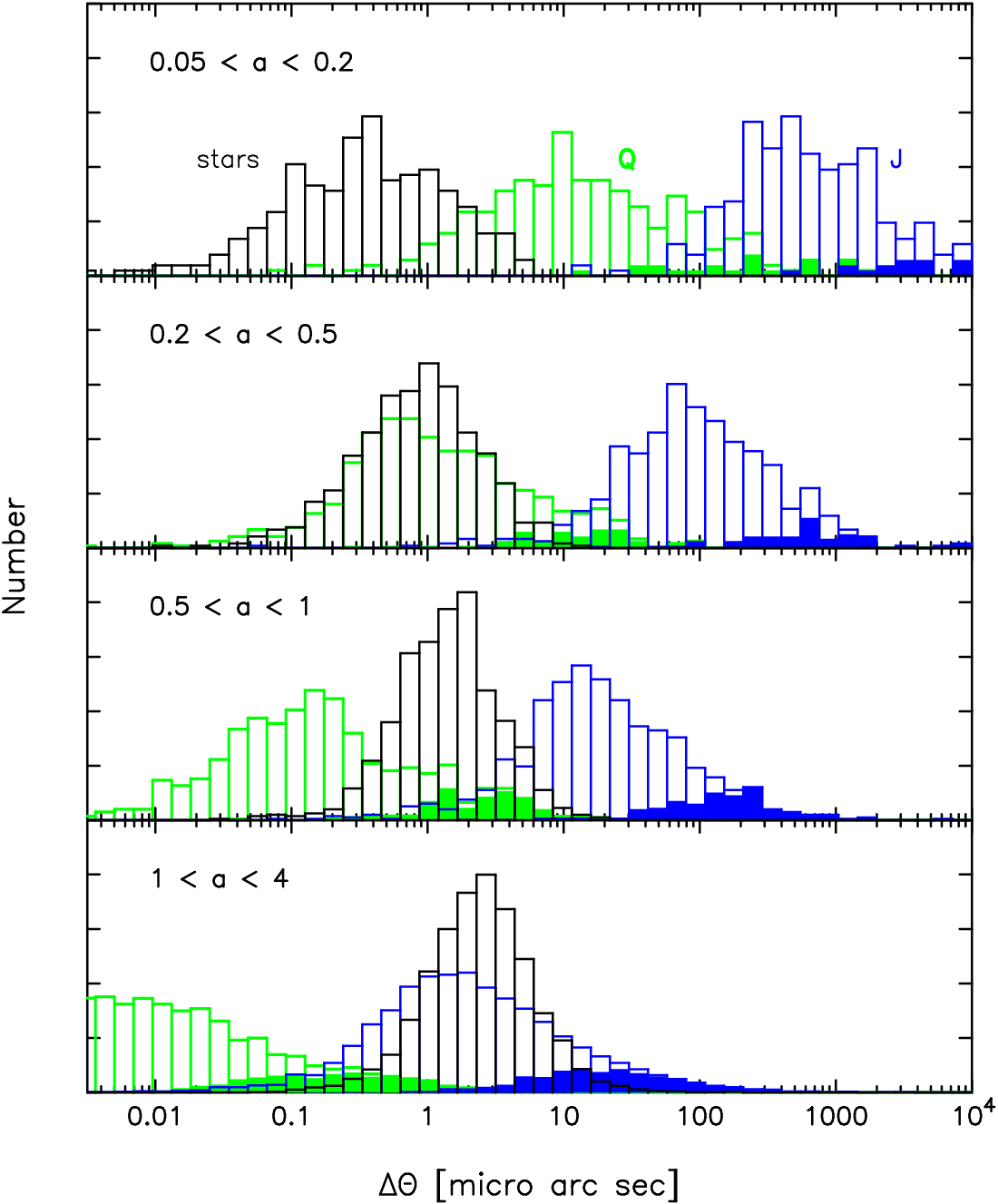}
\includegraphics[width=8.5cm]{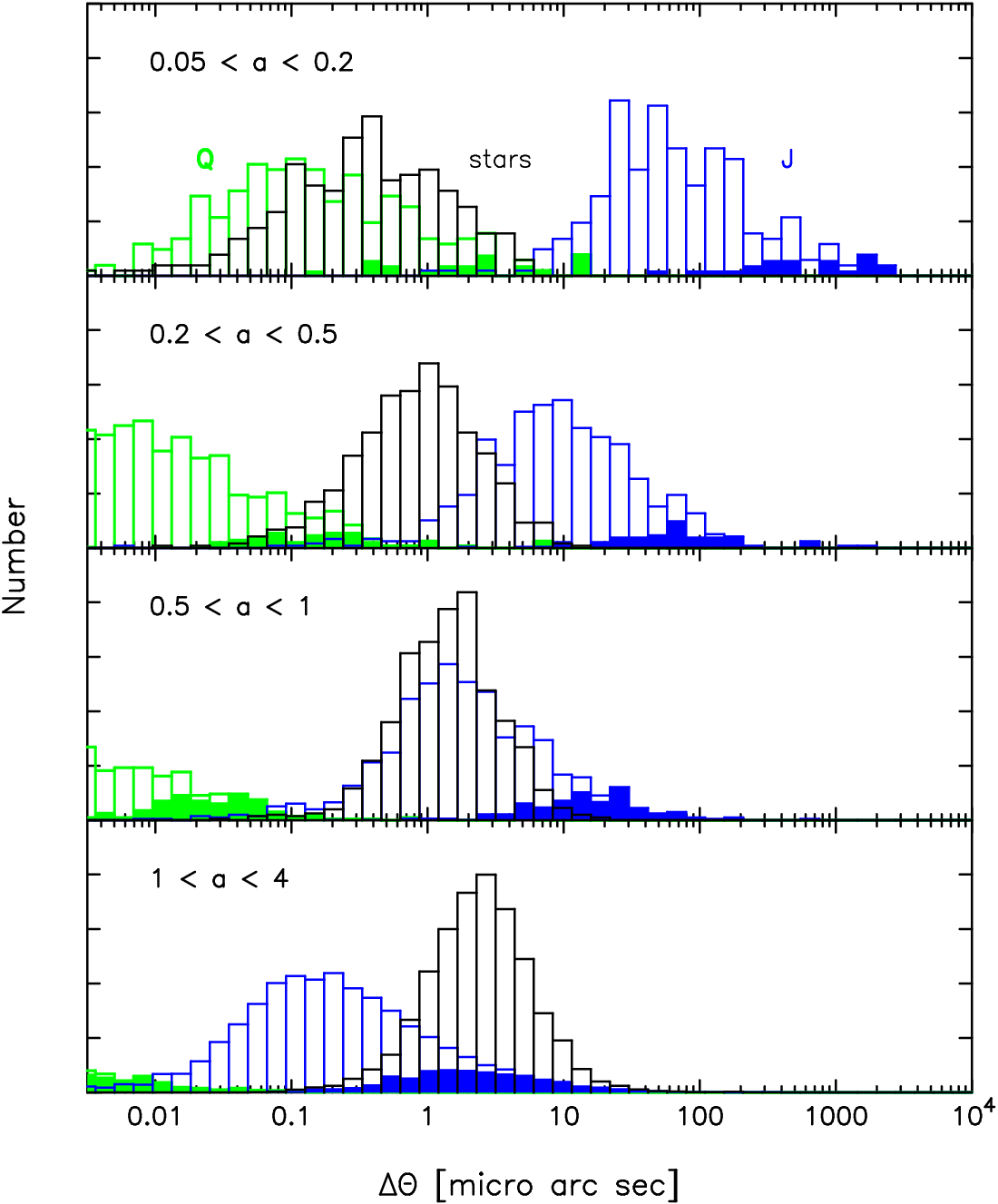}
\caption{\label{fig:stats} 
``Astrometric'' precessions of orbital planes over 10 years,
based on integrations of the model with 
($\gamma,\beta,\tilde M_\star$) $=$ ($2,0,30$).
Black histograms are from integrations in which the SBH spin was set to 
zero (although non-spin PN terms were included)
and show the effects of stellar perturbations.
Blue histograms are the predicted precessions due to frame-dragging alone
for the same stars;
filled bars are stars with $e\ge 0.9$.
Green histograms show the predicted precessions due to the SBH
quadrupole moment alone, again with high-$e$ orbits indicated.
The assumed value of the SBH spin is $\chi=1$ in the top frames
and $\chi=0.1$ in the bottom frames.
}
\end{figure}

Fig.~\ref{fig:all_bin} shows the results of a 
comprehensive set of integrations using different models for the
stellar cluster (Table~ 1).
Especially when the SBH spin is low ($\chi=0.1$), stellar perturbations can
dominate the signal due  to frame-dragging down to very small distances from the SBH,
e.g. $\sim 0.2$ mpc for $\tilde M_\star=100$,
corresponding to orbital periods of $\sim 0.1$ yr.

We define the ``astrometric precessions'' 
$\Delta\Theta \equiv (a/D)\Delta\theta$ where $D=8.0$ kpc is the 
distance to the Galactic center \cite{Will-08}.
$\Delta\Theta$ is roughly the angular displacement of the orbital axes,
as seen from the Earth
(ignoring projection effects).

In the following section we discuss how measurements of $\mathbf{J}$
and $Q$ might be feasible even in cases where the stellar perturbations
are significant.
Here, we assume that, in order to be useful for tests of GR, 
a star must satisfy two minimum conditions:
\begin{itemize}
\item[1] Its astrometric precession must exceed some minimum
threshold set by the detector.
\item[2] Its precession must be dominated by GR effects.
\end{itemize}
We call MS stars that satisfy both conditions ``detectable.''

We base our assumptions about the minimum observable angular changes on
the specifications for the planned instrument GRAVITY \cite{GRAVITY-09}.
GRAVITY will observe the Galactic
center three times a year (in April, July and September) and
the error in each astrometric data point will be 
$\delta\Theta \approx 10 \mu$as (F. Eisenhauer, private communication).
Since the precession is linear with respect to time,
the uncertainty in the measured astrometric precession after $n$ observations
is
\beq
\sigma_{\Delta\Theta} \approx 2\sqrt{3} 
\sqrt{n-1\over n(n+1)} \delta\Theta
\approx 35 \mu{\rm as}\ \sqrt{n-1\over n(n+1)}.
\eeq
For an elapsed time of $1(3)10$ yr, 
$\sigma_{\Delta\Theta} = 14(10)6.2\mu$as.

Fig.~\ref{fig:stats} plots the distribution of $\Delta\Theta$
values after 10 years due to star-star perturbations in one model,
($\gamma,\beta,\tilde M_\star$) $=$ ($2,0,30$).
Also plotted are the distributions that would arise
from frame-dragging and quadrupole torques alone,
$\Delta\Theta_{J,Q}=(a/D)\Delta\Omega_{J,Q}\sin i$
(cf. Eq.~\ref{eq:dtGR}), for the same stars.
Fig.~\ref{fig:stats} suggests that a clean separation of
frame-dragging and Newtonian precessions for most stars in this model
requires $a\lap 0.5$ mpc for $\chi=1$ and
$a\lap 0.2$ mpc for $\chi=0.1$.
For these values of $a$ the amplitude of $\Delta\Theta_J$ is
greater than $10 \mu$as for most stars, making them
accessible to astrometric monitoring.
High eccentricities, $e\gap 0.9$, allow these requirements
to be relaxed: eccentric orbits with $a$ as large as
$\sim 1 (0.5)$ mpc can 
produce measurable displacements that are signficantly
greater than those due to stellar perturbations.

Detecting the effects of the quadrupole moment in this model
would be considerably harder.
For $\chi=1$, the  quadrupole precessions separate cleanly
from the stellar perturbations only for $a\lap 0.1$ mpc,
or $a\lap 0.3$ for the most eccentric stars.

Regardless of the number of stars that satisfy conditions
1 and 2, an additional requirement is that 
\begin{itemize}
\item[3] the stars satisfying these conditions must 
constitute a large fraction of {\it all} stars in the region
being observed
\end{itemize}
since otherwise there is a large probability that the precession 
of a randomly-chosen star will be dominated by non-GR effects.

We calculated the average number of detectable stars,
$\langle N_{\rm detect}\rangle$,  in each of our models
after expressing conditions 1 and 2 in the forms
\begin{itemize}
\item[1] $\Delta\Theta > \sigma_{\Delta\Theta}$
\item[2] $\Delta\Theta_{J,Q} > \Delta\Theta_{95}$
\end{itemize}
where $\Delta\Theta_{95}$ is the upper edge of the 95\% 
confidence interval of the distribution of stellar 
perturbations.
Observational intervals of $\Delta t =(1,3,10)$ yr were considered.

\begin{figure*}
\includegraphics[width=14.cm]{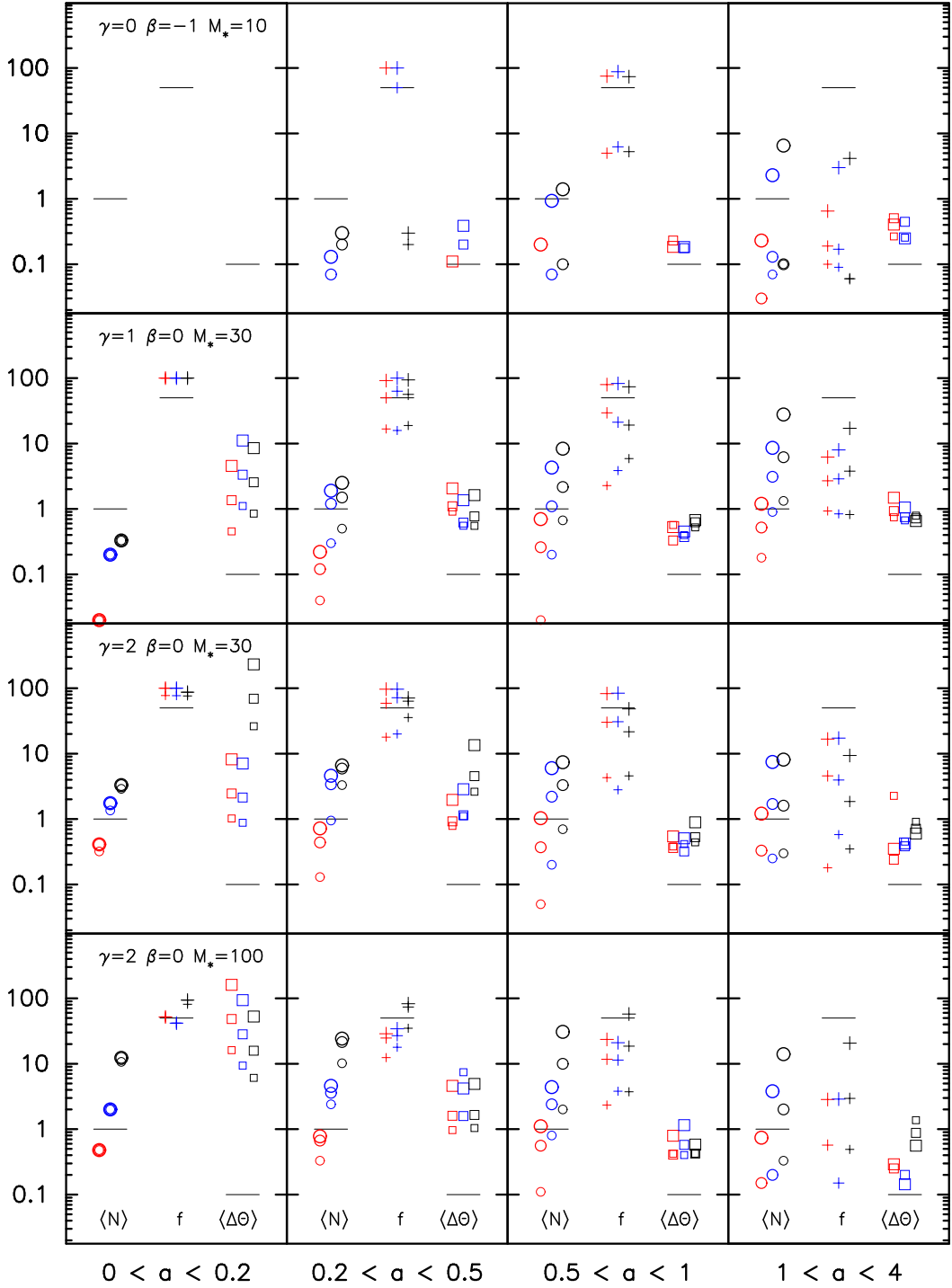}
\caption{\label{fig:all_stats_J} 
Detectability of frame-dragging in four models for the stellar cluster.
Results are displayed in four bins of semi-major axis $a$
(in mpc).
In each panel, circles denote $\langle N_{\rm detect}\rangle$,
the average number of stars with detectable precessions, as defined
in the text;
$+$ symbols denote the percentage of stars that are detectable
in that bin;
and $\square$ symbols denote the average precession angle of the detectable
stars in the bin, expressed in units of $10^{-4}$ arc min.
In each group of similar symbols, the left (red), middle (blue)
and right (black) symbols are for models with ${\cal R}=(1,0.1,0)$
respectively, while the integration time is indicated by the size
of the symbol: 1 yr (smallest), 3 yr, and 10 yr (biggest).
The tick marks indicate $\langle N_{\rm detect}\rangle=1$,
$f=50\%$, and $\langle\Delta\Theta\rangle=10\mu$as.
}
\end{figure*}

\begin{figure*}
\includegraphics[width=14.cm]{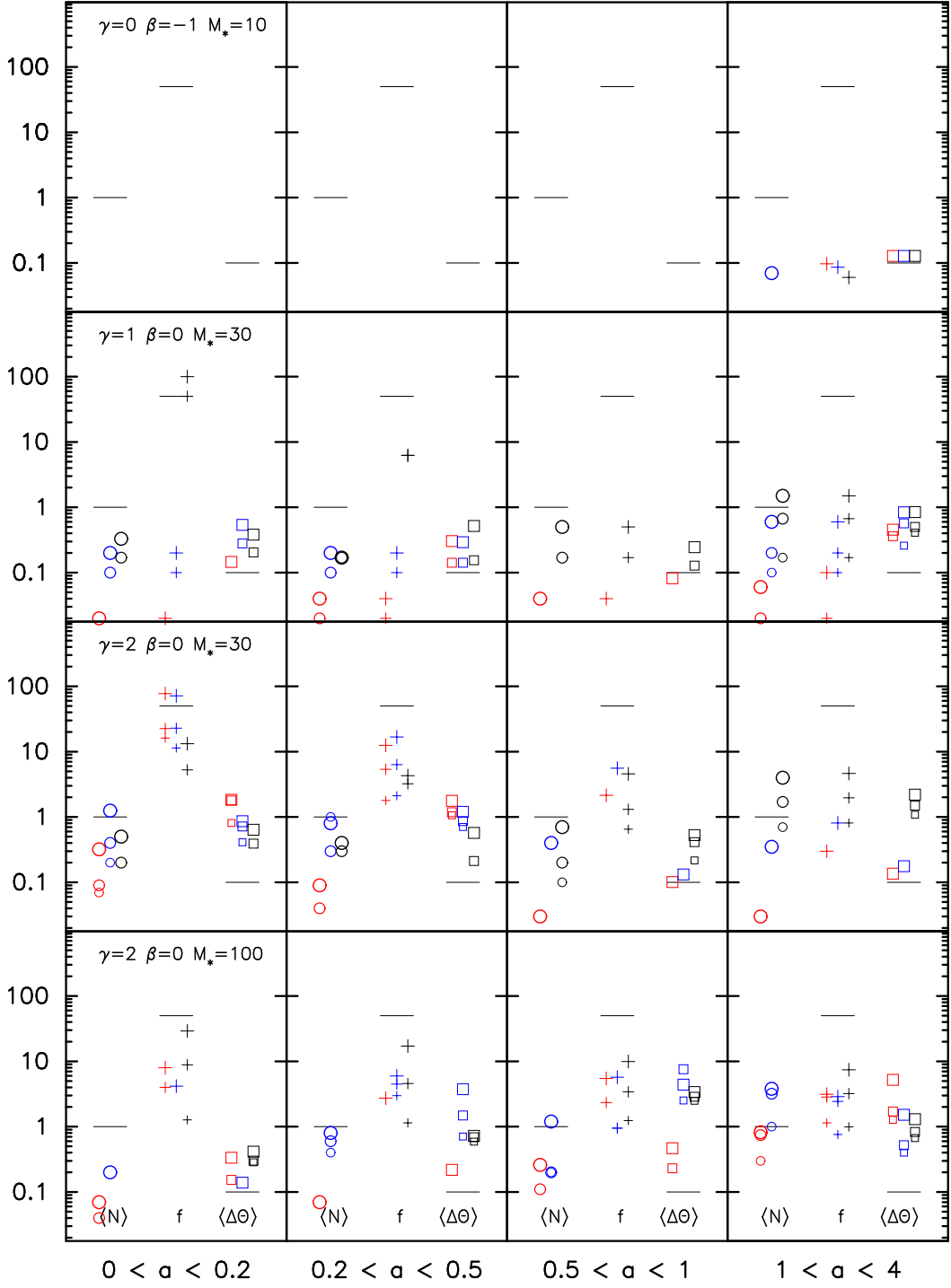}
\caption{\label{fig:all_stats_Q} 
Detectability of quadrupole precession. 
All symbols are defined as in Fig.~\ref{fig:all_stats_J}.}
\end{figure*}

The results are shown in Figs.~\ref{fig:all_stats_J} and
\ref{fig:all_stats_Q} for four models of the stellar cluster,
and in four bins of semi-major axis.
The figures also show $f$, the ratio of $\langle N_{\rm detect}\rangle$
to the total number of stars in each bin,
and $\langle\Delta\Omega\rangle$, the average value of
the ``astrometric precession'' for the detectable stars. 
$\langle N_{\rm detect}\rangle\gap 1$ means that at least one star
would be expected to be present that satisfies the two
detectability criteria; if in addition $f$  is large,
such stars constitute a large fraction of all stars in
the same radial bin.

The figures illustrate the tradeoff that occurs between
the number of detectable stars at a given radius,
and the certainty that a single star at that radius
is responding to GR effects rather than to stellar perturbations.
For example, going across the second column in 
Fig.~\ref{fig:all_stats_J} (i.e. $\gamma=1,\beta=0,\tilde M_\star=30$),
$\langle N\rangle$ increases steadily with increasing
$a$, but $f$ behaves oppositely, dropping below $10\%$ in most
cases for $a\gap 1$ mpc.

With regard to frame dragging Fig.~\ref{fig:all_stats_J} 
suggests the following.
\begin{itemize}
\item In models with low central densities, $\gamma=(0,1)$,
detection of frame-dragging precession may be feasible after 
$\Delta t\gap 3$ yr for orbits with
$0.2 \lap a/\mathrm{mpc}\lap 1 $.
At smaller radii the number of stars is too small;
at larger radii the noise from stellar perturbations is too great.
\item In models with a steep density profile,
$\gamma=2$, detection of frame dragging is feasible at 
$a\le 0.2$ mpc after $\Delta t = 1$ yr in most of the models;
at $a\le 0.5$ mpc after $\Delta t = 3$ yr;
and at $a\le 1$ mpc after $\Delta t = 10$ yr.
The exceptions are models with a large fraction of
stellar BHs (${\cal R}$=1) in which the stellar perturbations
always dominate.
\end{itemize}

With regard to quadrupole precession,
Fig.~\ref{fig:all_stats_Q} suggests that 
$\langle N_{\rm detect}\rangle>1$ occurs
in tandem with large ($\gap 50\%$) $f$ only for rather narrow sets
of parameters, e.g.
$\gamma\approx 2,
\tilde M \approx 30,
a\lap 0.2 \mathrm{mpc} ,
{\cal R}\lap 0.1,
\Delta t\gap 10$ yr.
Detecting the effects of the quadrupole torque above noise
from the stellar perturbations is apparently
only feasible if the stellar cluster is rather finely tuned:
there must be a substantial number of stars very close to the
SBH, $r\lap 0.2$ mpc, but at the same time a small number of
stellar remnants so that the stellar perturbations do
not dominate.

In practice, the quadrupole-induced precession may be large
compared with that due to stellar perturbations, but still
small compared with precession due to frame dragging, 
making it difficult to see in the data.
As shown in Fig.~\ref{fig:at},
$t_Q<t_J$ only holds at very small radii, $r\lap 0.05$ mpc,
unless eccentricities are large.

\section{\label{sec:discuss} Discussion}
\subsection{Compensating for stellar perturbations}

The numerical experiments described above were designed to
elucidate the extent to which ``noise'' from
stellar perturbations can mask the signal from GR spin effects
at the Galactic center.
In situations where the stellar perturbations are present but
not dominant, one would like to be able to detect the perturbations
and remove their effects from the data.
Here we outline one approach to that problem.

Suppose that one observes a set of stars at the discrete
times $t_j, j=1,\ldots,N_{\mathrm{obs}}$.
Denoting the stars by index $k=1,\ldots,N_{\mathrm{orb}}$,
the $N_\mathrm{orb}$ orbital solutions are required
to satisfy the equations 
\begin{subequations}
\begin{eqnarray}
\frac{\mathrm{d}\mathbf{L}_k}{\mathrm{d}t} &=&
\boldsymbol{\omega}_k\times\mathbf{L}_{k}(t_{j}) +
\mathbf{r}(t_j)\times \mathbf{F}_{N}(t_{j},\mathbf{r})\,,\\
\boldsymbol{\omega}_{k} &=& P_k^{-1}
\left(A_{J,k}-A_{Q,k}\cos i_{k}\right)\hat{\mathbf{J}}
\label{eq:w}
\end{eqnarray}
\label{eq:torque}
\end{subequations}
Here $\mathbf{F}_{N}(t,\mathbf{r})$ is the perturbing force field due to
all the stars and stellar remnants, and $\mathbf{r}_k(t)$ is the
position of the $k$th star at time $t_j$.
$\mathbf{F}_{N}$ depends explicitly on the time, to the extent
that the orbits of the stars that produce the perturbing
force are changing.

We first consider the orbital solutions for each observed star separately.
In their most general form, equations (\ref{eq:torque}) are under-determined
($3N_{\mathrm{obs}}$ equations with $3+3N_{\mathrm{obs}}$ unknowns).
To make progress, it is necessary to introduce approximations. 
On the short time scales of relevance, the contribution to the
total gravitational potential from the stars can 
be considered (1) fixed in time and (2) not strongly varying
in space.
For instance, a spatially uniform $\mathbf{F}$ is the lowest
order term in a multipole expansion, and one could add,
if necessary, dipole and quadrupole terms, etc.

As an example, we consider a stellar perturbation that is modeled as
\beq
\Phi_\star(x,y,z) = 2\pi G\rho_t\left(A_xx^2 + A_yy^2 + A_zz^2\right).
\eeq
This is the potential due to a homogeneous triaxial cluster centered
on the SBH and aligned with the coordinate axes; the dimensionless
quantities ($A_x,A_y,A_z$) are determined by the axis
ratios of the ellipsoid \cite{Chandra-69}.
Proceeding as in the derivation of Eq.~(\ref{eq:dOmega}), one
can derive the (Newtonian) orbit-averaged effect of this perturbing
potential on the orientations of orbital planes.
For instance, in the case of low-eccentricity orbits, one finds
\cite{SS-00}
\begin{subequations}
\begin{eqnarray}
\frac{\mathrm{d}\mathbf{L}}{\mathrm{d}t} &=& 
\mathbf{T}_{N}\,,\\
\mathbf{T}_N &=& \frac{4\pi G\rho_t a^2}{L^2}\left(
\begin{array}{c}
(A_z-A_x)L_yL_z\\
(A_y-A_z)L_xL_z\\
(A_x-A_y)L_xL_y
\end{array}
\right)
\end{eqnarray}
\end{subequations}
with $L$ constant. Similar expressions can be derived for
orbits of arbitrary eccentricity \cite{SS-00}.

In practice, one would simultaneously fit all the parameters
appearing in Eq.~(\ref{eq:torque}) to the observed positions and/or
velocities.
In the case of the model potential just assumed, this would 
amount to introducing 9 additional parameters 
to the 6 required for a Keplerian solution, i.e.
$\boldsymbol{\omega}_{k}$;
the perturbing density $\rho_t$ and its two axis ratios;
and the three direction cosines that define the orientation of the ellipsoid.
A data set with $N_{\mathrm{obs}}\ge5$ 
is then formally well-determined. 
Note however that $N_{\mathrm{obs}}\gg5$
is required to obtain a full solution with a reasonable goodness-of-fit.
Repeating this fitting procedure for the $N_{\mathrm{orb}}$ orbits
will then allow one to estimate the orbit-to-orbit variation in 
the parameters that define the perturbation.
If the latter is large, the model can be made more general by
the addition of extra parameters.

This fitting procedure will yield an estimate of $\boldsymbol{\omega}_{k}$
for the $N_{\mathrm{orb}}$ orbits that is independent of any assumptions
about the underlying physics of the GR precession, apart from the assumption
that the precession rate is constant. Given enough orbits, the empirically
determined values $\boldsymbol{\omega}_{k}$ can be correlated with
the orbital properties to test in an almost assumption-free way non-standard
theories of gravity.
The no-hair conjecture
can be tested explicitly be substituting into equation (\ref{eq:w})
the GR expressions (\ref{eq:tJtQ}):
\begin{eqnarray}
A_{J,k} & = & 4\pi\chi\varrho_{k}^{-3/2}\,,\nonumber \\
A_{Q,k} & = & 3\pi\chi^{2}\varrho_{k}^{-2}\end{eqnarray}
with $\varrho$ the penetration parameter defined above.
 The torque equation (\ref{eq:torque}) can then be written as 
\begin{eqnarray}
&&\frac{\mathrm{d}\mathbf{L}_{k}}{\mathrm{d}t}=4\pi\left(\varrho_{k}^{-3/2}/P_{k}\right)\nonumber \\
&&\left[1+\frac{3}{4}\varrho_{k}^{-1/2}\boldsymbol{\chi}\cdot(\mathbf{L}_{k}/L_{k})\right]\left(\boldsymbol{\chi}\times\mathbf{L}_{k}\right)+\mathbf{T}_{N,k}\,\label{eq:dLk}
\end{eqnarray}
The second (small) term in the square brackets ($\varrho_{k}\gg1$) reflects
the relative contribution of the quadrupole moment to the precession.
The same fitting procedure outlined above for the general case can
be carried out here, with the difference that the vector $\boldsymbol{\chi}$,
which replaces $\boldsymbol{\omega}_{k}$, is common to all the orbits,
and therefore a simultaneous fitting of the entire set of $N_{\mathrm{orb}}$
orbits will improve the power of the orbital solutions. The comparison
of the best fit values of $\boldsymbol{\chi}$ from individual orbits
to the one from the simultaneous global fit can help assess the robustness
of the result. 

An initial test of the no-hair conjecture could be obtained by repeating
the fit once with, and once without the quadrupole term, to check
whether the quadrupole term indeed improves the fit. A slightly more
discriminating test would be to introduce an extra free pre-factor
$f_{Q}$ to the quadrupole term, and see whether the best fit yields
$f_{Q}\sim1$. More sophisticated tests would require assuming specific
alternative functional forms for the quadrupole term.

\subsection{Independent constraints on the spin}

The foregoing assumes that the mass $\mh$ of the Milky
Way SBH is a known quantity, i.e. that it need not
be treated as a free parameter when fitting the astrometric data
to Eq.~(\ref{eq:torque}).
Determinations of $\mh$ are based on the motion
of stars on spatial scales of $10^1-10^3$ mpc 
\cite{Ghez-08,Gillessen-09a,Schoedel-09},
outside the region where GR spin effects are detectable.

In the same way, independent measurements of the magnitude and
direction of the SBH's spin $\mathbf{J}$
could remove as many as three additional
parameters from Eq.~(\ref{eq:torque}), reducing the uncertainty
on estimates of the quadrupole moment.

A number of  approaches to the determination of  
$\mathbf{J}$  are being pursued.
Sgr A$^*$, the compact source of radio, infrared and X-radiation at 
the center of the Milky Way, exhibits variability at IR and
X-ray wavelengths on time scales as short as $20-30$ minutes that can 
be interpreted as emission from ``hot spots'' in orbits just
outside the SBH event horizon
\cite{Genzel-03b,Yusef-06,Eckart-06,Marrone-08}.
If this interpretation is correct, 
$\chi\gap 0.3-0.5$ is required in order that the orbital period
at the innermost stable orbit be as short as the observed
rise times for the flaring events \cite{Genzel-03b,Belanger-06,Meyer-06}.
Very long baseline interferometry at sub-mm wavelengths
has the capacity to resolve the structure of the plasma
surrounding the SBH on angular scales of $10-100$ $\mu$as,
comparable to $r_g$ \cite{Doeleman-08,Fish-09}.
Such observations can constrain the spin via comparison
with theoretical accretion disk models, or by 
detection of variability on spatial scales of $\sim r_g$
\cite{Doeleman-09,Gammie-09,Broderick-09}.
The properties of the X-ray polarization from the inner accretion
disk are also predicted to be spin-dependent \cite{SK-09}.

Constraints on $\mathbf{J}$ derived
from observations like these will be highly model-dependent, 
requiring assumptions about the nature of the emission, the geometry
and physical state of the emitting gas (accretion vs. outflow or jets),
etc.
It is not clear at present whether the systematic
uncertainties of these independent spin measurements will be larger or smaller 
than the uncertainties associated with astrometric estimates of the spin;
the latter are due both to the stellar perturbations modelled here, 
and also to the difficulties associated with measurement of 
proper motions of faint stars in crowded fields
\cite{Fritz-09}.
Additional, if uncertain, information on the SBH spin
can be used to improve the global orbital solutions by including
it as prior probabilities in Bayesian best-fit methods for large-dimensional 
parameter spaces, such as the Markov Chain Monte Carlo method
\cite{Metro-53,Hastings-70}.

\section{\label{sec:conclude}Conclusions}

\noindent
1. The spin and quadrupole moment of the Galactic center
supermassive black hole (SBH) can in principle be measured
by observing the precession of the orbital planes 
of stars in the inner milliparsec (mpc) \cite{Will-08}.
However, gravitational interactions between stars in this region
are likely to induce orbital precession of the same approximate
amplitude as the precession due to frame dragging.
The stellar perturbations manifest themselves as coherent
torques over short time scales, 
mimicking general relativistic (GR) precession.

\noindent
2. The number of stars and stellar remnants (e.g. stellar-mass
black holes, BHs) in this region is uncertain, but small enough
($\sim 10^0 - 10^3$) that full $N$-body simulations are feasible.
A regularized post-Newtonian $N$-body algorithm is presented that includes 
the lowest-order spin-orbit and quadrupole-orbit terms.

\noindent
3. Assuming near-maximal spin for the Milky Way SBH,
detection of frame-dragging precession may be feasible after a few years'
monitoring with an instrument like GRAVITY \cite{GRAVITY-09} 
for orbits in the radial range between $\sim 0.2$ mpc and $\sim 1$ mpc.
At smaller radii the number of stars is too small, while
at larger radii the star-star and star-remnant perturbations
dominate GR effects.
In models where the number of stellar BHs is comparable
to the number of observable stars, GR effects are almost always
swamped by perturbations from the remnants.

\noindent
4. Quadrupole-induced precession stands out clearly from
stellar perturbations only in a narrow class of models
for the nuclear star cluster,
having moderate to high central densities and a small BH fraction,
and only at radii $r\lap 0.2$ mpc.

\noindent
5. Because the orbit-averaged torques from stars are approximately
constant in magnitude over year-long time scales, it is possible
in principle to disentangle the effects of stellar perturbations
from those due to GR,
allowing tests of gravity even in the presence of stellar 
perturbations.

\begin{acknowledgments}
DM was supported in part by the National Science Foundation under  
grants no. AST 08-07910, 08-21141 and by the National Aeronautics and 
Space Administration under grant no. NNX-07AH15G.
TA was supported by ISF grant no. 928/06 and by ERC Starting Grant
202996.
CMW was supported in part by the National Science Foundation under  
grant no. PHY 06-52448 and the National Aeronautics and Space 
Administration under grant no. NNG-06GI60G.  
CMW is grateful for the  hospitality of the Institut d'Astrophysique 
de Paris where parts of  this research were carried out. 
We thank H. Bartko, F. Eisenhauer and S. Noble for useful discussions.
\end{acknowledgments}

\bibliography{ms}

\end{document}